\documentclass[12pt,a4paper]{article}
\usepackage{graphicx}
\usepackage{fancyhdr}
\usepackage{setspace}
\usepackage{color}

\usepackage{nfontdef}

\newcommand{\ignore}[1]{}

\newcommand{\vek}[1]{\mathchoice{\displaystyle\boldsymbol#1}
{\textstyle\boldsymbol#1}{\scriptstyle\boldsymbol#1}
{\scriptscriptstyle\boldsymbol#1}}
\newcommand{\mat}[1]{\mathchoice{\displaystyle\mathbf#1}
{\textstyle\mathbf#1}{\scriptstyle\mathbf#1}
{\scriptscriptstyle\mathbf#1}}

\begin{document}
\title{Modeling of Degradation Processes in Historical Mortars}
\author{J. S\'{y}kora\footnote{Corresponding author. Tel. +420 224 354 495; fax: +420 224 310 775. E-mail address: jan.sykora.1@fsv.cvut.cz}}
\maketitle
\centerline{$^*$Department of Mechanics, Faculty of Civil Engineering}
\centerline{Czech Technical University in Prague}
\centerline{Th\'{a}kurova 7, 166 29 Prague, Czech Republic}

\vspace{0.2in}
\centerline{Dedicated to Professor Zden\v{e}k Bittnar in ocassion of his 70th birthday.}
\vspace{0.2in}

\begin{abstract}
The aim of presented paper is modeling of degradation processes in
historical mortars exposed to moisture impact during freezing.
Internal damage caused by ice crystallization in pores is one of
the most important factors limiting the service life of historical
structures. Coupling the transport processes with the mechanical
part will allow us to address the impact of moisture on the
durability, strength and stiffness of mortars. This should be
accomplished with the help of a complex thermo-hygro-mechanical
model representing one of the prime objectives of this work. The
proposed formulation is based on the extension of the classical
poroelasticity models with the damage mechanics. An example of
two-dimensional moisture transport in the environment with
temperature below freezing point is presented to support the
theoretical derivations.

\end{abstract}

\section*{Nomenclature}
\label{sec:nomen}

\renewcommand{\arraystretch}{1.1}
\begin{table}[h!]
\begin{center}
\begin{tabular}{|p{1.5cm}p{13cm}|}
\hline
$a$                     & water absorption coefficient, $\mathrm{[kgm^{-2}s^{-0.5}]}$ \\
$\vek{b}$               & body force, $\mathrm{[Nm^{-3}]}$ \\
$b$                     & Biot's coefficient, $\mathrm{[-]}$ \\
$b_{\mathrm{tcs}}$      & thermal conductivity supplement, $\mathrm{[-]}$ \\
$b_{\varphi}$           & approximation factor, $\mathrm{[-]}$ \\
$\mat{C}$               & discretized capacity matrix \\
$c_{\mathrm{i}}$        & specific heat capacity of ice, $\mathrm{[Jkg^{-1}K^{-1}]}$  \\
$c_{\mathrm{s}}$        & specific heat capacity of solid matrix, $\mathrm{[Jkg^{-1}K^{-1}]}$ \\
$c_{\mathrm{l}}$        & specific heat capacity of liquid water, $\mathrm{[Jkg^{-1}K^{-1}]}$ \\
$\mat{D}_{e}$           & elastic stiffness matrix, $\mathrm{[Pa]}$  \\
$D_{\mathrm{l}}$        & capillary transport coefficient, $\mathrm{[m^{2}s^{-1}]}$  \\
$D_{\varphi}$           & liquid conduction coefficient, $\mathrm{[kgm^{-1}s^{-1}]}$  \\
$d_{w}$                 & damage parameter, $\mathrm{[-]}$  \\
$E$                     & Young's modulus, $\mathrm{[Pa]}$  \\
$\vek{F}$               & prescribed fluxes \\
$f_{t}$                 & tensile strength, $\mathrm{[Pa]}$  \\
$H$                     & total enthalpy of porous material, $\mathrm{[Jm^{-3}]}$ \\
$h_{\mathrm{i}}$        & specific melting enthalpy of ice, $\mathrm{[Jkg^{-1}]}$ \\
$h_{\mathrm{v}}$        & latent heat of phase exchange, $\mathrm{[Jkg^{-1}]}$ \\
$\mat{K}$               & discretized conductivity matrix \\
$K$                     & bulk modulus of porous material, $\mathrm{[Pa]}$  \\
$K_{\mathrm{s}}$        & bulk modulus of solid matrix, $\mathrm{[Pa]}$  \\
$l_{\mathrm{intl}}$     & internal length, $\mathrm{[m]}$  \\
$M_{w}$                 & molar mass of water, $\mathrm{[kgmol^{-1}]}$ \\
$\vek{n}$               & unit normal vector, $\mathrm{[-]}$\\
\end{tabular}
\end{center}
\end{table}

\begin{table}[h!]
\begin{center}
\begin{tabular}{|p{1.5cm}p{13cm}|}
$n$                         & total porosity, $\mathrm{[-]}$ \\
$p_{\mathrm{a}}$            & atmospheric pressure, $\mathrm{[Pa]}$ \\
$p_{\mathrm{l}}$            & liquid pressure, $\mathrm{[Pa]}$ \\
$p_{\mathrm{p}}$            & average pore pressure, $\mathrm{[Pa]}$ \\
$p_{\mathrm{sat}}$          & saturation vapor pressure, $\mathrm{[Pa]}$ \\
$\vek{q}_{\mathrm{h}}$      & heat flux, $\mathrm{[Wm^{-2}]}$ \\
$\vek{q}_{\mathrm{l}}$      & liquid transport flux, $\mathrm{[kgm^{-2}s^{-1}]}$ \\
$\vek{q}_{\mathrm{v}}$      & water vapor flux, $\mathrm{[kgm^{-2}s^{-1}]}$ \\
$R$                         & gas constant, $\mathrm{[Jmol^{-1}K^{-1}]}$ \\
$r$                         & pore radius, $\mathrm{[m]}$ \\
$\vek{r}$                   & nodal values \\
$r_{\mathrm{ar}}$           & layer of adsorbed water, $\mathrm{[m]}$ \\
$r_{\mathrm{cr}}$           & critical pore radius, $\mathrm{[m]}$ \\
$r_{\mathrm{ir}}$           & curvature radius of ice crystal, $\mathrm{[m]}$ \\
$S_{h}$                     & heat source, $\mathrm{[Wm^{-3}]}$ \\
$S_{w}$                     & moisture source, $\mathrm{[kgm^{-3}s^{-1}]}$ \\
$t$                         & time, $\mathrm{[s]}$ \\
$\vek{u}$                   & displacement vector, $\mathrm{[m]}$ \\
$w$                         & total water content, $\mathrm{[kgm^{-3}]}$ \\
$w_{\mathrm{80}}$           & water content at $0.8\,\mathrm{[-]}$ relative humidity, $\mathrm{[kgm^{-3}]}$ \\
$w_{\mathrm{i}}$            & content of ice, $\mathrm{[kgm^{-3}]}$ \\
$w_{f}$                     & free water saturation, $\mathrm{[kgm^{-3}]}$ \\
$\alpha$                    & thermal expansion coefficient, $\mathrm{[K^{-1}]}$  \\
$\alpha_{\mathrm{h}}$       & heat transfer coefficient, $\mathrm{[Wm^{-2}K^{-1}]}$ \\
$\alpha_{\mathrm{swr}}$     & short wave absorption coefficient, $\mathrm{[-]}$ \\
$\beta_{\mathrm{v}}$        & water vapor transfer coefficient, $\mathrm{[kgm^{-2}s^{-1}Pa^{-1}]}$ \\
$\Gamma$                    & boundary \\
$\gamma$                    & generalized midpoint integration parameter, $\mathrm{[-]}$ \\
$\gamma_{\mathrm{li}}$      & liquid/ice surface tension, $\mathrm{[Nm^{-1}]}$ \\
$\Delta s_{\mathrm{m}}$     & melting entropy, $\mathrm{[PaK^{-1}]}$\\
$\delta$                    & vapor diffusion coefficient in air, $\mathrm{[kgm^{-1}s^{-1}Pa^{-1}]}$ \\
$\delta_{\mathrm{v}}$       & water vapor permeability of porous material, $\mathrm{[kgm^{-1}s^{-1}Pa^{-1}]}$ \\
$\varepsilon_{\mathrm{0}}$  & equivalent strain at elastic limit, $\mathrm{[-]}$  \\
$\varepsilon_{\mathrm{eq}}$ & equivalent strain, $\mathrm{[-]}$  \\
$\varepsilon_{f}$           & equivalent strain at critical crack opening, $\mathrm{[-]}$  \\
$\theta$                    & temperature, $\mathrm{[^{\circ}C]}$ \\
$\lambda$                   & thermal conductivity of moist porous material, $\mathrm{[Wm^{-1}K^{-1}]}$ \\
$\lambda_{\mathrm{0}}$      & thermal conductivity of dry porous material, $\mathrm{[Wm^{-1}K^{-1}]}$ \\
\end{tabular}
\end{center}
\end{table}

\begin{table}[h!]
\begin{center}
\begin{tabular}{|p{1.5cm}p{13cm}|}
$\mu$                       & water vapor diffusion resistance factor, $\mathrm{[-]}$ \\
$\nu$                       & Poisson's ratio, $\mathrm{[-]}$ \\
$\rho_{\mathrm{s}}$         & bulk density, $\mathrm{[kgm^{-3}]}$ \\
$\vek{\sigma}$              & total stress, $\mathrm{[Pa]}$ \\
$\vek{\sigma}'$             & effective stress, $\mathrm{[Pa]}$ \\
$\varphi$                   & relative humidity, $\mathrm{[-]}$ \\
$\chi$                      & local pressure on the frozen pore walls, $\mathrm{[Pa]}$ \\
$\psi$                      & cumulative volume of pores, $\mathrm{[-]}$ \\
$\Omega$                    & domain \\
                            & \\
Subscripts                  & \\
ext                         & exterior \\
i                           & ice \\
in                          & initial \\
int                         & interior \\
l                           & liquid water \\
p                           & pore \\
r                           & radius \\
s                           & solid \\
v                           & water vapor \\
\hline
\end{tabular}
\end{center}
\end{table}

\section{Introduction}
\label{sec:intro}

Understanding the hydro-thermo-mechanical behavior of building
materials exposed to weather conditions is the first step toward
avoiding deterioration of structures in general and historical
ones in particular, as high moisture content in building material
and its phase changes are often a cause of internal damage.
Because of variable climatic conditions, the moisture gradients
induce mechanical stresses in the porous material. These stresses
mostly develop due to the growth of ice crystals through the pore
structure. Therefore, there is a strong need for investigating the
influence of moisture on the mechanical material behavior, which
leads to numerical and experimental coupling of mechanical and
thermo-hygro phenomena.

In the literature, the above described problem is addressed from
several perspectives. The first group of publications is focused
on the description of the coupled heat and moisture transport
reflecting the moisture migration under the conditions of the ice
crystal formation in the pores, $2$-D and $3$-D aspects and
different moisture/heat sources, such as wind driven rain, solar
short and long wave radiation etc., see
~\cite{Kong:2011:EB,Kunzel:1996:IJHMT,Tan:2011:CRT}. An extensive
overview of various transport models is available
in~\cite{Lewis:1999,Cerny:2002}. While models for transport
processes have been developed during several decades, the theory
of ice crystallization in the pores has emerged only recently,
~\cite{Scherer:1993:JNS,Scherer:1999:CCR,Sun:2010:CCR}. The
authors established relations between physical state of porous
system and pore pressures. The physical conditions of ice
formation process are described by thermodynamic balance equation
between ice, liquid water and solid matrix. Finally, the
mechanical response of porous media subjected to the frost action
was studied by several authors
~\cite{Coussy:2008:CCR,Wardeh:2008:CBM,Zuber:2000:CCR}. On the one
hand, the poroelasticity formulation based on Biot's continuum
model was adopted. It is an efficient method for elastic modeling
of porous system, which is subjected to the pressure of the fluid.
On the other hand, a novel micromechanics approach was introduced
to analyze the creation of micro-cracks in the microstructure
during freezing process ~\cite{Koster:2010:RILEM,Liu:2011:CRT}.
These results predict effective mechanical and transport
properties at microscopic level and can be utilized as an input
for multi-scale analysis of porous media.

As a preamble, our goal is to quantify the internal damage caused
by the ice crystallization pressure in historical mortars. In
particular, a critical point in a restoration works is frequent
applications of lime mortars for preserving compatibility with the
historical materials. However, lime mortars are very porous, their
mechanical strength and durability are mostly very low
~\cite{Nunes:2012:EM,Slizkova:2010:ROUD}, thus the development of
a lime mortar with improved internal hydrophobicity and associated
improved resistance against damage due to the effects of ice
crystallization is inevitable. To address this issue with respect
to its complexity, an analysis combining both experimental work
and numerical simulations has to be done. Nevertheless, the
numerical methodology developed within this work can be utilized
to simulate the response of any porous material subjected to the
frost action.

A theoretical formulation of the problem is presented in
Section~\ref{sec:mm}, followed by numerical calculations in
Section~\ref{sec:ne}. Section~\ref{sec:InfPSD} then investigates
the influence of pore size distribution on the evolution of damage
parameter. The essential findings are summarized in
Section~\ref{sec:con}.

\section{Material model}
\label{sec:mm}

\begin{figure} [h!]
\begin{center}
\begin{tabular}{c}
\includegraphics*[width=130mm,keepaspectratio]{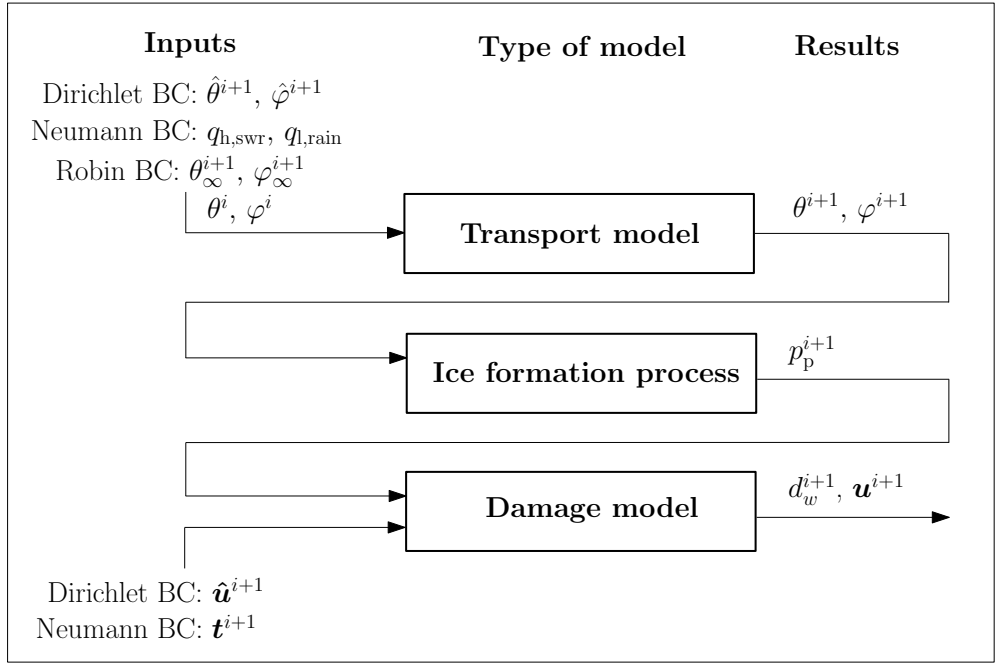}\\
\end{tabular}
\end{center}
\caption{Algorithmic framework of proposed model}
\label{fig:framework}
\end{figure}

The problem of porous system subjected to ice crystallization can
be divided into three physical phenomena - heat and moisture
transport, ice formation process and evolution of damage caused by
pore pressure. Using the thermodynamics, poromechanics and damage
mechanics, we propose here the concept of multi-phase constitutive
model based on the assumption of the uncoupled system in the sense
of numerical analysis, see Fig.~\ref{fig:framework}. These models
are characterized by combining different physical or mechanical
models (in space and time) in order to accurately describe
structural response of deteriorating infrastructure over time. The
general framework of the proposed model was primarily inspired by
the work published
in~\cite{Bazant:2002:JEM,Biot:2002:JAP,Coussy:2008:CCR,Gawin:2003:CMAME,Kunzel:1996:IJHMT,Scherer:1993:JNS,Wardeh:2008:CBM,Zuber:2000:CCR}.

In the presented work, the porous material is treated as
multi-phase medium consisting of solid matrix, liquid water, water
vapor and ice. The mathematical formulation consists of three
governing equations representing the conservations of energy
Eq.~(\ref{eq:TE03}), mass Eq.~(\ref{eq:TE04}) and linear momentum
Eq.~(\ref{eq:linmom}). The chosen primary unknowns are temperature
$\theta\,\mathrm{[^{\circ} C]}$, moisture $\varphi\,\mathrm{[-]}$
and displacement of solid matrix $\vek{u}\,\mathrm{[m]}$.


The present section derives the governing equations of the
analytical model. After theoretical formulation, a proper
numerical time and space integration scheme is introduced to
convert the proposed governing equations into a fully discrete
form. Some details on the numerical implementation are also
available in~\cite{Kucerova:2013:AMC,Sykora:2013:AMC}.

\subsection{Transport model}
We use the diffusion model by K\"{u}nzel,
see~\cite{Kunzel:1996:IJHMT}, which is based on Krischer's
concept~\cite{Cerny:2002}.
K\"{u}nzel neglected the liquid water and water vapor convection
driven by gravity and total pressure as well as enthalpy changes
due to liquid flow and choose relative humidity $\varphi$ as the
only moisture driving force. The water vapor diffusion is then
described by Fick's law written as
\begin{equation}
\vek{q}_{\mathrm{v}}=-\delta_{\mathrm{v}}\vek{\nabla}\left(\varphi
p_{\mathrm{sat}}\right),\label{eq:TFlux-1}
\end{equation}
where $\vek{q}_{\mathrm{v}}\,\mathrm{[kgm^{-2}s^{-1}]}$ is the
water vapor flux,
$\delta_{\mathrm{v}}\,\mathrm{[kgm^{-1}s^{-1}Pa^{-1}]}$ is the
water vapor permeability of a porous material and
$p_{\mathrm{sat}}=p_{\mathrm{sat}}(\theta)\,\mathrm{[Pa]}$ is the
saturation water vapor pressure being exponentially dependent on
temperature. The transport of liquid water is assumed in the form
of surface diffusion in an absorbed layer and capillary flow
typically represented by Kelvin's law
\begin{equation}
\vek{q}_{\mathrm{l}} =
-D_{\varphi}\vek{\nabla}\varphi,\label{eq:TFlux-2}
\end{equation}
where $\vek{q}_{\mathrm{l}}\,\mathrm{[kgm^{-2}s^{-1}]}$ is the
flux of liquid water,
$D_{\varphi}=D_{\mathrm{l}}\,(\mathrm{d}w/\mathrm{d}\varphi)\,\mathrm{[kgm^{-1}s^{-1}]}$
is the liquid conductivity and
$D_{\mathrm{l}}\,\mathrm{[m^{2}s^{-1}]}$ is the capillary
transport coefficient, $\mathrm{d}w/\mathrm{d}\varphi$ is the
derivative of water retention function. The Fourier law is then
used to express the heat flux
$\vek{q}_{\mathrm{h}}\,\mathrm{[Wm^{-2}]}$ as
\begin{equation}
\vek{q}_{\mathrm{h}} =
-\lambda\vek{\nabla}\theta,\label{eq:TFlux-3}
\end{equation}
where $\lambda\,\mathrm{[Wm^{-1}K^{-1}]}$ is the thermal
conductivity and $\theta\,\mathrm{[^{\circ} C]}$ is the local
temperature. Introducing the above constitutive equations into
energy and mass conservation equations we finally get resulting
set of differential equations for the description of heat and
moisture transfer expressed in terms of temperature and relative
humidity as
\begin{itemize}
\item the energy balance equation
\begin{equation}
\frac{\partial H}{\partial\theta}\frac{\partial\theta}{\partial t}
 =  \vek{\nabla}\cdot[\lambda\vek{\nabla} \theta]+h_{\mathrm{v}}\vek{\nabla}\cdot
[\delta_{\mathrm{v}}\vek{\nabla}\{\varphi
p_{\mathrm{sat}}(\theta)\}], \label{eq:TE03}
\end{equation}
\item the conservation of mass equation
\begin{equation}
\frac{\mathrm{d}w}{\mathrm{d}\varphi}\frac{\partial
\varphi}{\partial t}  =  \vek{\nabla}\cdot
[D_{\varphi}\vek{\nabla}\varphi]+\vek{\nabla}\cdot
[\delta_{\mathrm{v}}\vek{\nabla}\{\varphi
p_{\mathrm{sat}}(\theta)\}] \, , \label{eq:TE04}
\end{equation}
\end{itemize}
The transport coefficients defining the material behavior are
nonlinear functions of the temperature, moisture and material
properties. We briefly recall their particular
expressions~\cite{Kunzel:1996:IJHMT}: 

\begin{itemize}
\item $w$ - total water content $\mathrm{[kgm^{-3}]}$,
    \begin{equation}
    w = w_{f}\frac{(b_{\varphi}-1)\varphi}{b_{\varphi}-\varphi}, \label{eq:DTC1}
    \end{equation}
    where $w_{f}\,\mathrm{[kgm^{-3}]}$ is the free water saturation and $b_{\varphi}\,\mathrm{[-]}$ is the
    approximation factor, which must always be greater than one. It
    can be determined from the equilibrium water content ($w_{80}$) at
    $0.8$ [-] relative humidity by substituting the corresponding
    numerical values in Eq.~(\ref{eq:DTC1}).
    Fig.~\ref{fig:matpar}(a) shows an example of variation of water content as a function of relative humidity.

\item $\delta_{\mathrm{v}}$ - water vapor permeability
    $\mathrm{[kgm^{-1}s^{-1}Pa^{-1}]}$,
    \begin{equation}
    \delta_{\mathrm{v}} = \frac{\delta}{\mu},
    \end{equation}
    where $\mu\,\mathrm{[-]}$ is the water vapor diffusion resistance factor and $\delta\,\mathrm{[kgm^{-1}s^{-1}Pa^{-1}]}$ is the vapor diffusion
    coefficient in air given by
    \begin{equation}
    \delta = \frac{2.306\cdot{}10^{-5 }\,p_{\mathrm{a}}}{R_{\mathrm{v}}\,(\theta+273.15)\,p}\left(\frac{\theta+273.15}{273.15}\right )^{1.81},
    \end{equation}
    with $p$ set equal to atmospheric pressure $p_{a}=101325$ [Pa] and
    $R_{\mathrm{v}}$ = $R$/$M_{w} = 461.5$ [$\mathrm{Jkg^{-1}K^{-1}}$]; $R$
    is the gas constant ($8314.41$ [Jmol$^{-1}$K$^{-1}$]) and $M_w$
    is the molar mass of water ($18.01528$ [kgmol$^{-1}$]).
    An example of variation of water vapor permeability as a function of temperature is seen in Fig.~\ref{fig:matpar}(b).

\begin{figure} [t!]
\begin{center}
\begin{tabular}{cc}
\includegraphics*[width=75mm,keepaspectratio]{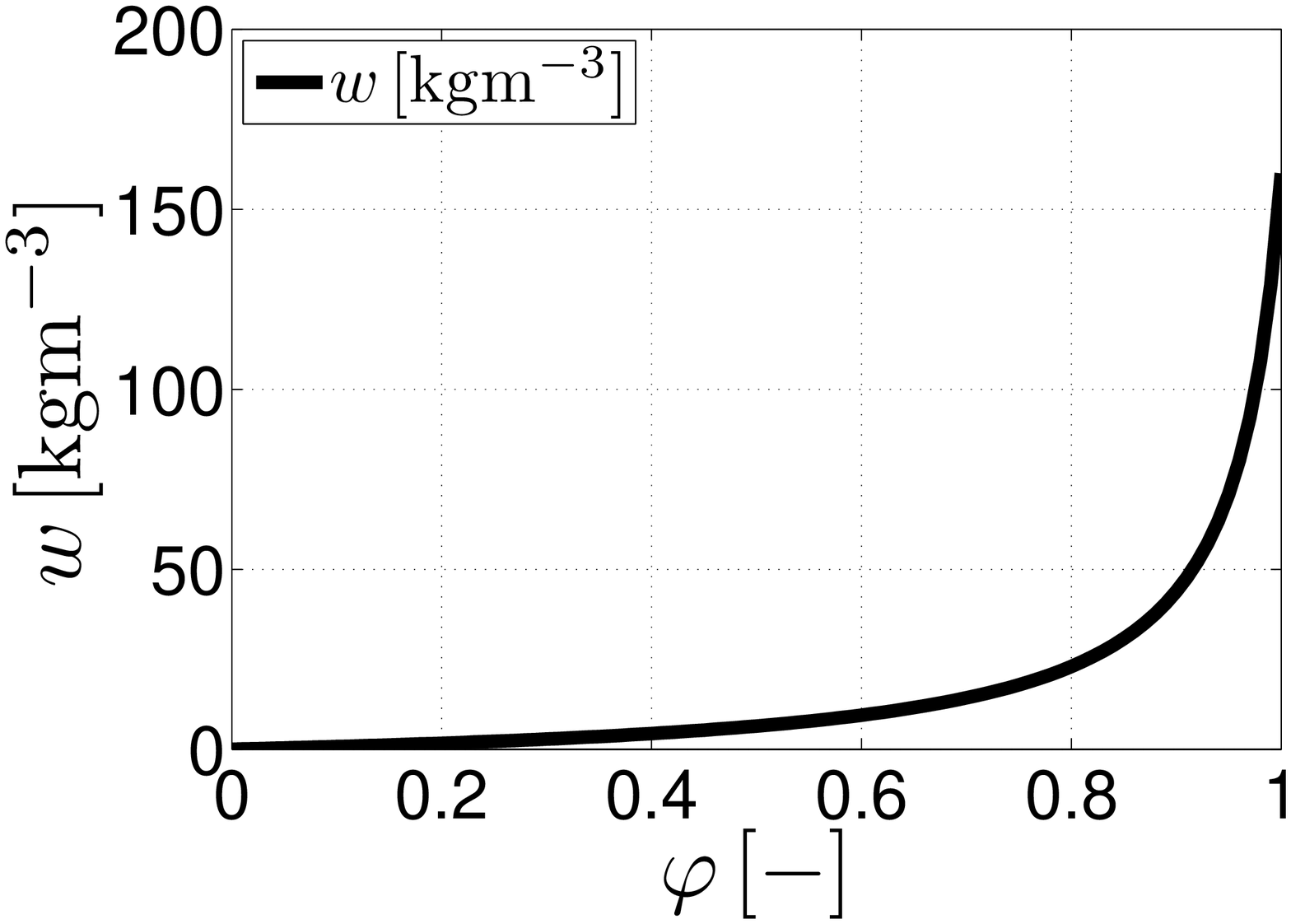}&
\includegraphics*[width=75mm,keepaspectratio]{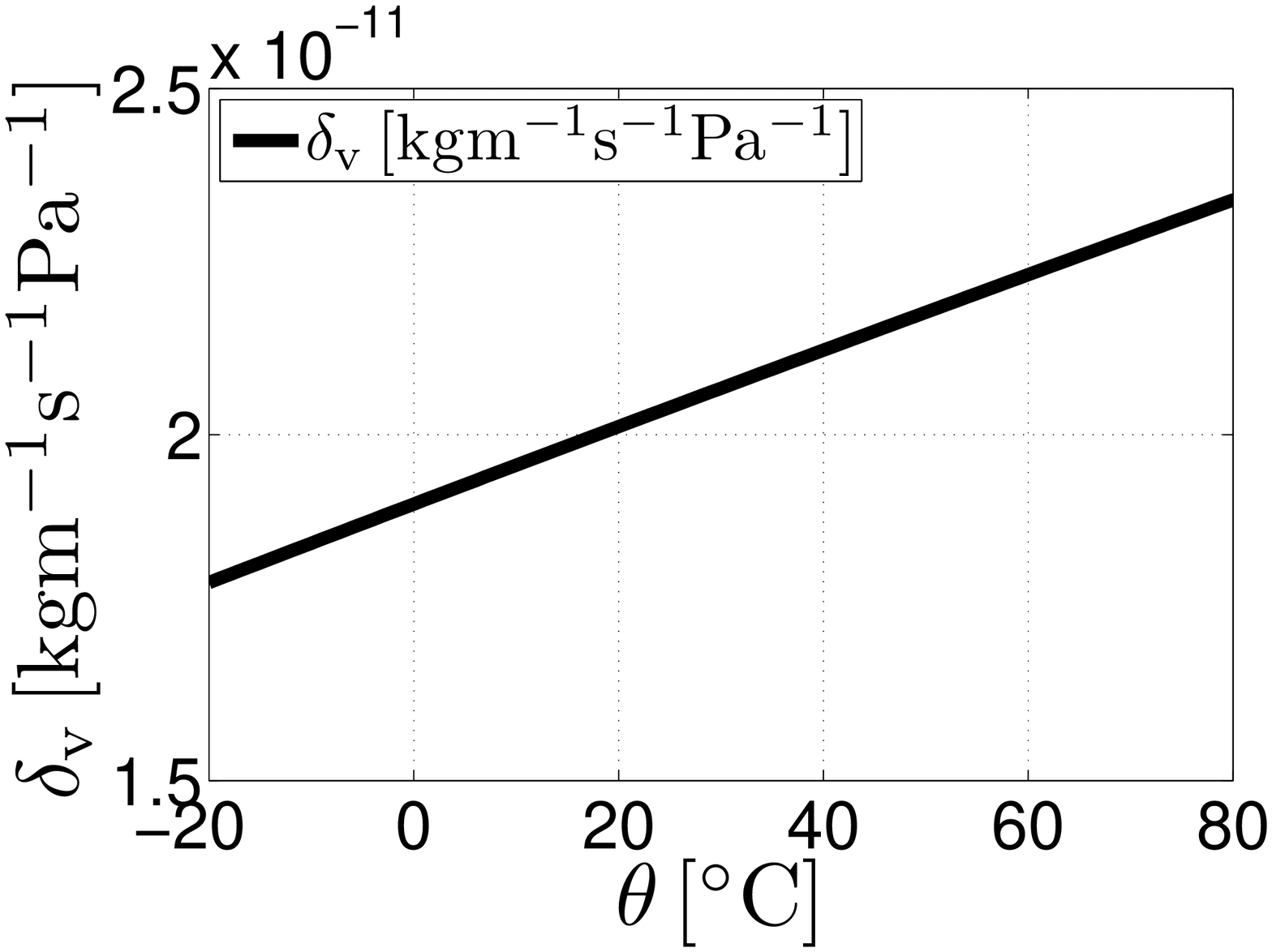}\\
(a)&(b) \\
\includegraphics*[width=75mm,keepaspectratio]{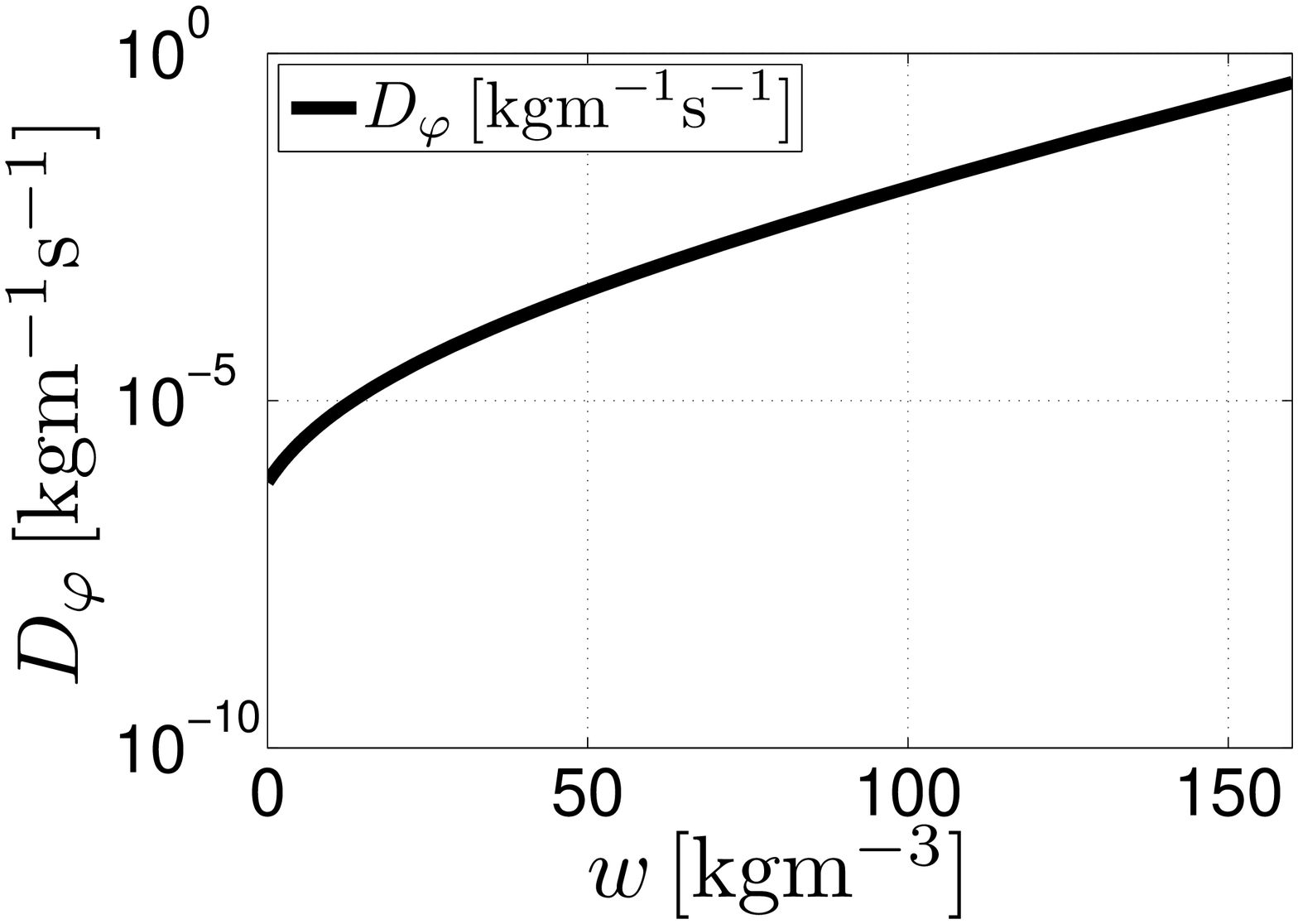}&
\includegraphics*[width=75mm,keepaspectratio]{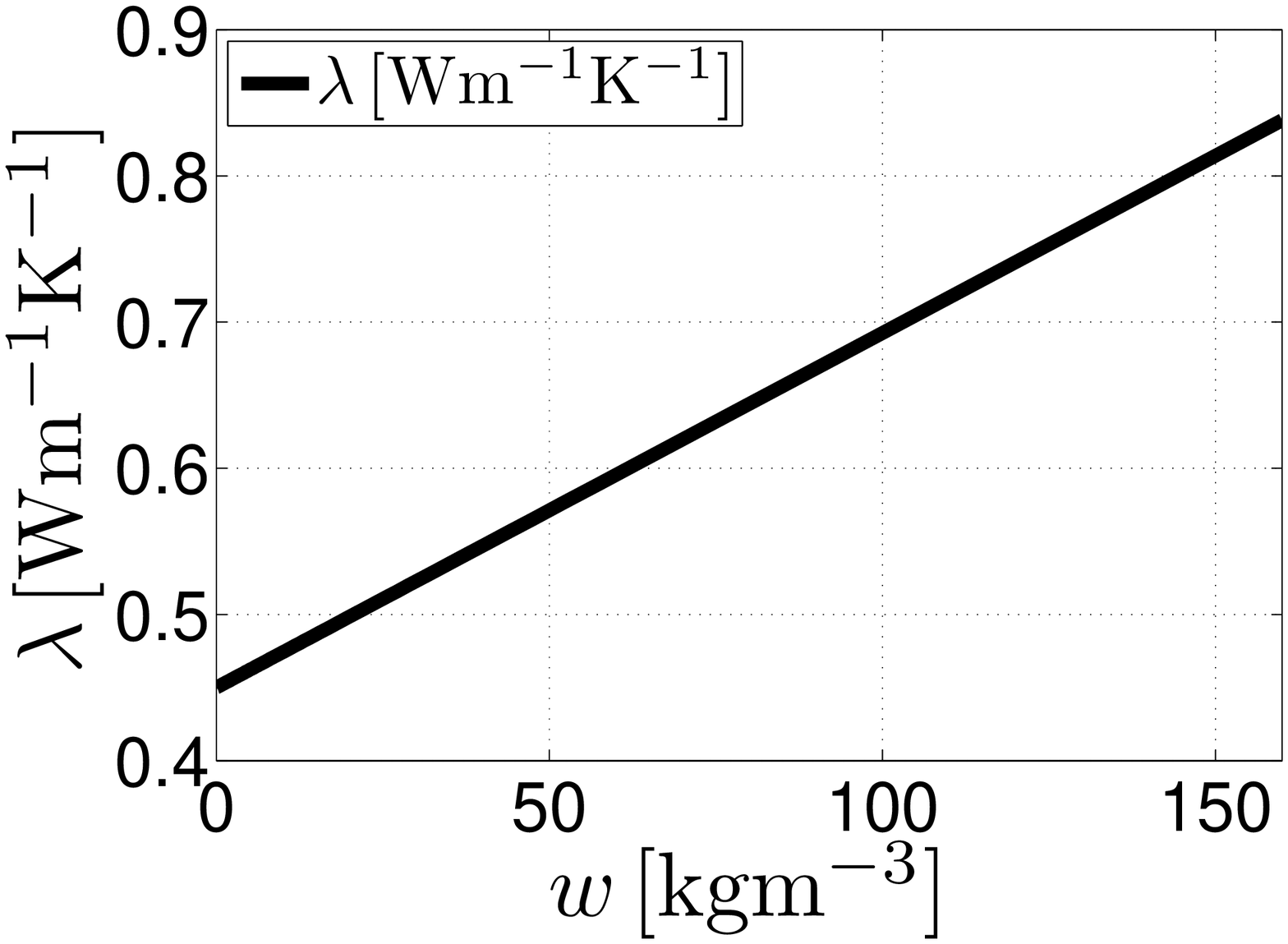}\\
(c)&(d) \\
\end{tabular}
\end{center}
\caption{(a) Variation of water content as a function of relative
humidity, (b) variation of water vapor permeability as a function
temperature, (c) variation of liquid conductivity as a function of
water content, (d) variation of thermal conductivity as a function
of water content} \label{fig:matpar}
\end{figure}

\item $D_{\varphi}$ - liquid conduction coefficient
    $\mathrm{[kgm^{-1}s^{-1}]}$,
    \begin{equation}
    D_{\varphi}=D_{\mathrm{l}}\frac{\mathrm{d}w}{\mathrm{d}\varphi},
    \end{equation}
    where $D_{\mathrm{l}}\,\mathrm{[m^{2}s^{-1}]}$ is the capillary transport coefficient given by
    \begin{equation}
    D_{\mathrm{l}}=3.8\left(\frac{a}{w_{f}}\right
    )^{2}\cdot{}10^{3w/(w_{f}-1)},
    \end{equation}
    An example of variation of
    liquid conductivity $D_\varphi\,\mathrm{[kgm^{-1}s^{-1}]}$ as a function of water content
    is plotted in Fig.~\ref{fig:matpar}(c).

\item $\lambda$ - thermal conductivity $\mathrm{[Wm^{-1}K^{-1}]}$,
    \begin{equation}
    \lambda=\lambda_{0}\left(1+\frac{b_{\mathrm{tcs}}w}{\rho_{\mathrm{s}}}\right),
    \end{equation}
    where $\lambda_{0}\,\mathrm{[Wm^{-1}K^{-1}]}$ is the thermal conductivity of dry building
    material, $\rho_{\mathrm{s}}\,\mathrm{[kgm^{-3}]}$ is the bulk density and $b_{\mathrm{tcs}}\,\mathrm{[-]}$ is
    the thermal conductivity supplement. An example of variation of thermal
    conductivity as a function of water content is shown in Fig.~\ref{fig:matpar}(d).

\item $p_{\mathrm{sat}}$ - water vapor saturation pressure
    $\mathrm{[Pa]}$,
    \begin{equation}
    p_{\mathrm{sat}}=611\,\exp\left(\frac{a\,\theta}{\theta_{0}+\theta}\right),
    \end{equation}
    where
    \begin{equation}
    \begin{array}{ccc}
    a=22.44 & \theta_{0}=272.44\,[^{\circ}\mathrm{C}] &
    \theta<0\,[^{\circ}\mathrm{C}]\\[0.5cm]
    a=17.08 & \theta_{0}=234.18\,[^{\circ}\mathrm{C}] &
    \theta\geq0\,[^{\circ}\mathrm{C}]
    \end{array}
    \end{equation}

\item $h_{\mathrm{v}}$ - evaporation enthalpy of water
$\mathrm{[Jkg^{-1}]}$
    \begin{equation}
    h_{\mathrm{v}} = 2.5008 \cdot{} 10^{6} \left(\frac{273.15}{\theta}\right)^{(0.167 +
    3.67 \cdot{} 10^{-4}\theta )}.
    \end{equation}

\item $H$ - total enthalpy of porous material $[\mathrm{Jm}^{-3}]$
    \begin{equation}
    H = \rho_{\mathrm{s}} c_{\mathrm{s}} \theta +
    \left[(w-w_{\mathrm{i}})c_{\mathrm{l}}+w_\mathrm{i}c_\mathrm{i}-h_{\mathrm{i}}\frac{\mathrm{d}
    w_{\mathrm{i}}}{\mathrm{d}
    \theta}\right]\theta,\label{eq:enthalpy}
    \end{equation}
    where $\rho_{\mathrm{s}}\,\mathrm{[kgm^{-3}]}$ is the bulk density, $c_{\mathrm{i}}\,\mathrm{[Jkg^{-1}K^{-1}]}$ is the specific heat capacity of ice,
    $c_{\mathrm{s}}\,\mathrm{[Jkg^{-1}K^{-1}]}$ is the specific heat capacity of solid matrix, $c_{\mathrm{l}}\,\mathrm{[Jkg^{-1}K^{-1}]}$ is the
    specific heat capacity of liquid water and $h_{\mathrm{i}}\,\mathrm{[Jkg^{-1}]}$ is the specific melting enthalpy of ice, $w_{\mathrm{i}}\,\mathrm{[kgm^{-3}]}$ is the
    content of ice.

\end{itemize}

For the spatial discretization of the partial differential
equations, a finite element method is preferred here to the finite
volume technique. The discretized form of energy and moisture
balance equations then reads
\begin{equation}
\underbrace{\left[\begin{array}{cc}
\mat{K}_{\theta\theta}(\vek{r}_{\theta},\vek{r}_{\varphi}) &
\mat{K}_{\theta\varphi}(\vek{r}_{\theta},\vek{r}_{\varphi}) \\
\mat{K}_{\varphi\theta}(\vek{r}_{\theta},\vek{r}_{\varphi}) &
\mat{K}_{\varphi\varphi}(\vek{r}_{\varphi})\end{array}\right]}_{\mat{K}(\vek{r})}
\underbrace{\left\{\begin{array}{c} \vek{r}_{\theta} \\
\vek{r}_{\varphi}\end{array}\right\}}_{\vek{r}} +
\underbrace{\left[\begin{array}{cc}
\mat{C}_{\theta\theta}(\vek{r}_{\theta},\vek{r}_{\varphi}) &
0 \\
0 &
\mat{C}_{\varphi\varphi}(\vek{r}_{\varphi})\end{array}\right]}_{\mat{C}(\vek{r})}
\underbrace{\left\{\begin{array}{c} \frac{\mathrm{d}\vek{r}_{\theta}}{\mathrm{d}t} \\
\frac{\mathrm{d}\vek{r}_{\varphi}}{\mathrm{d}t}\end{array}\right\}}_{\dot{\vek{r}}}
= \underbrace{\left\{\begin{array}{c} \vek{F}_{\theta} \\
\vek{F}_{\varphi}\end{array}\right\}}_{\vek{F}}, \label{eq:disc}
\end{equation}
where $\mat{K}$ is the conductivity matrix, $\mat{C}$ is the
capacity matrix, $\vek{r}$ is the vector of nodal values, and
$\vek{F}$ is the vector of prescribed fluxes transformed into
nodes. For a detailed formulation of the matrices $\mat{K}$ and
$\mat{C}$ and the vector $\vek{F}$, we refer the interested reader
to~\cite{Sykora:2012:JCAM,Sykora:2013:AMC}.

The numerical solution of the system Eq.~(\ref{eq:disc}) is based
on a simple temporal finite difference discretization.  If we use
time steps $\Delta\, t$ and denote the quantities at time step $i$
with a corresponding superscript, the time-stepping equation is
\begin{equation}
\vek{r}^{i+1} = \vek{r}^{i} + \Delta t
[(1-\gamma)\dot{\vek{r}}^{i} + \gamma\dot{\vek{r}}^{i+1}],
\label{eq:ts}
\end{equation}
where $\gamma$ is a generalized midpoint integration rule
parameter. In the results presented in this paper the
Crank-Nicolson (trapezoidal rule) integration scheme with $\gamma
= 0.5$ was used.  Expressing $\dot{\vek{r}}^{i+1}$ from
Eq.~(\ref{eq:ts}) and substituting into the Eq.~(\ref{eq:disc}),
one obtains a system of non-linear equations:
\begin{equation}
  [\gamma \Delta t \mat{K}^{i+1}(\vek{r}^{i+1})  + \mat{C}^{i+1}(\vek{r}^{i+1})] \vek{r}^{i+1}
  = \gamma \Delta t \vek{F}^{i+1}  + \mat{C}^{i+1}(\vek{r}^{i+1}) [\vek{r}^{i}
  + \Delta t \{1-\gamma\} \dot{\vek{r}}^{i}],\label{eq:MS5}
\end{equation}
which can be solved by some iterative method such as
Newton-Raphson.

\subsection{Ice formation process}
The ice crystallization process in the porous system is described
by the penetration of liquid/ice interface from external surfaces
or large pores towards the unfrozen zones~\cite{Scherer:1993:JNS}.

\begin{figure} [t!]
\begin{center}
\begin{tabular}{cc}
\includegraphics*[width=76mm,keepaspectratio]{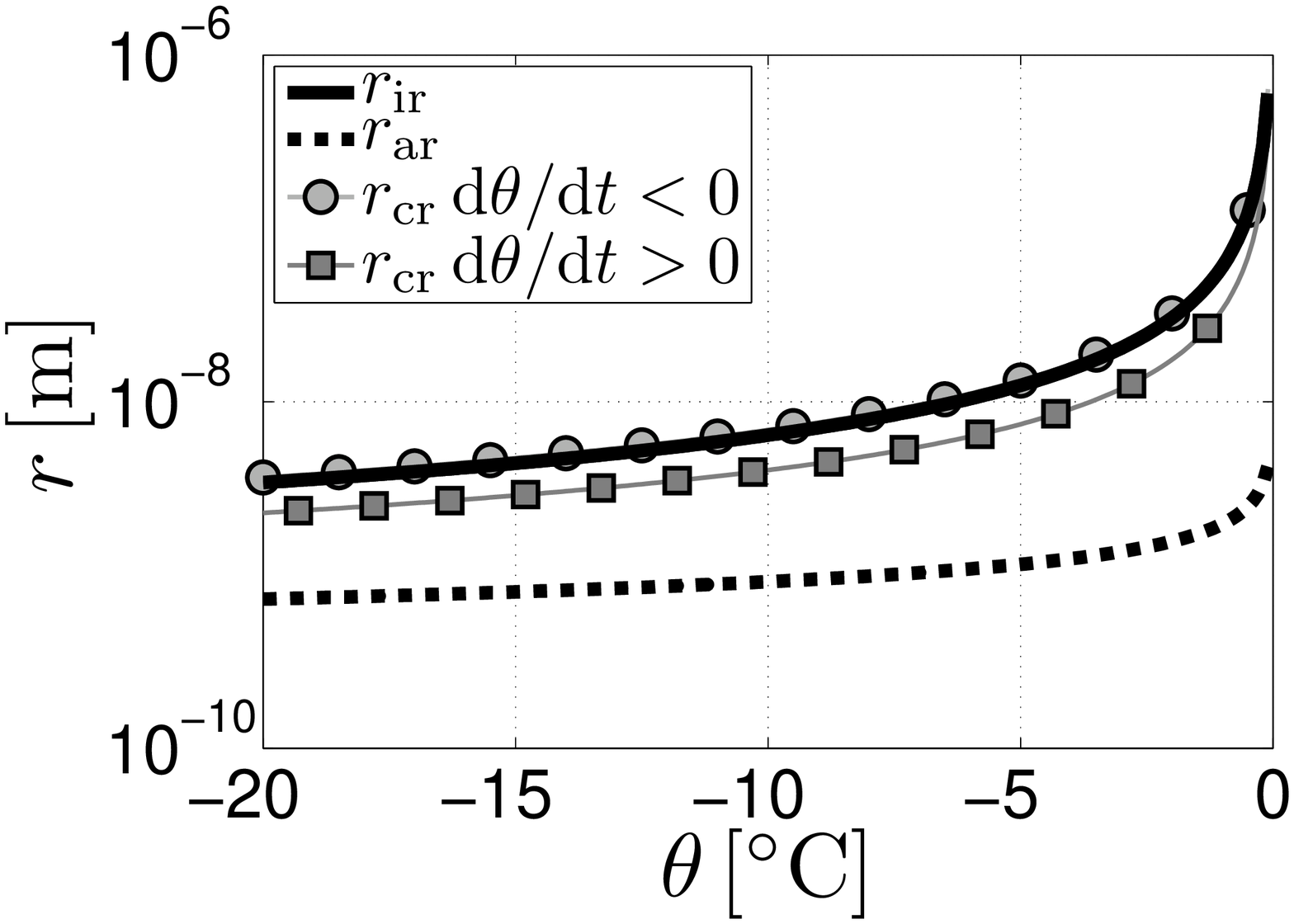}&
\includegraphics*[width=75mm,keepaspectratio]{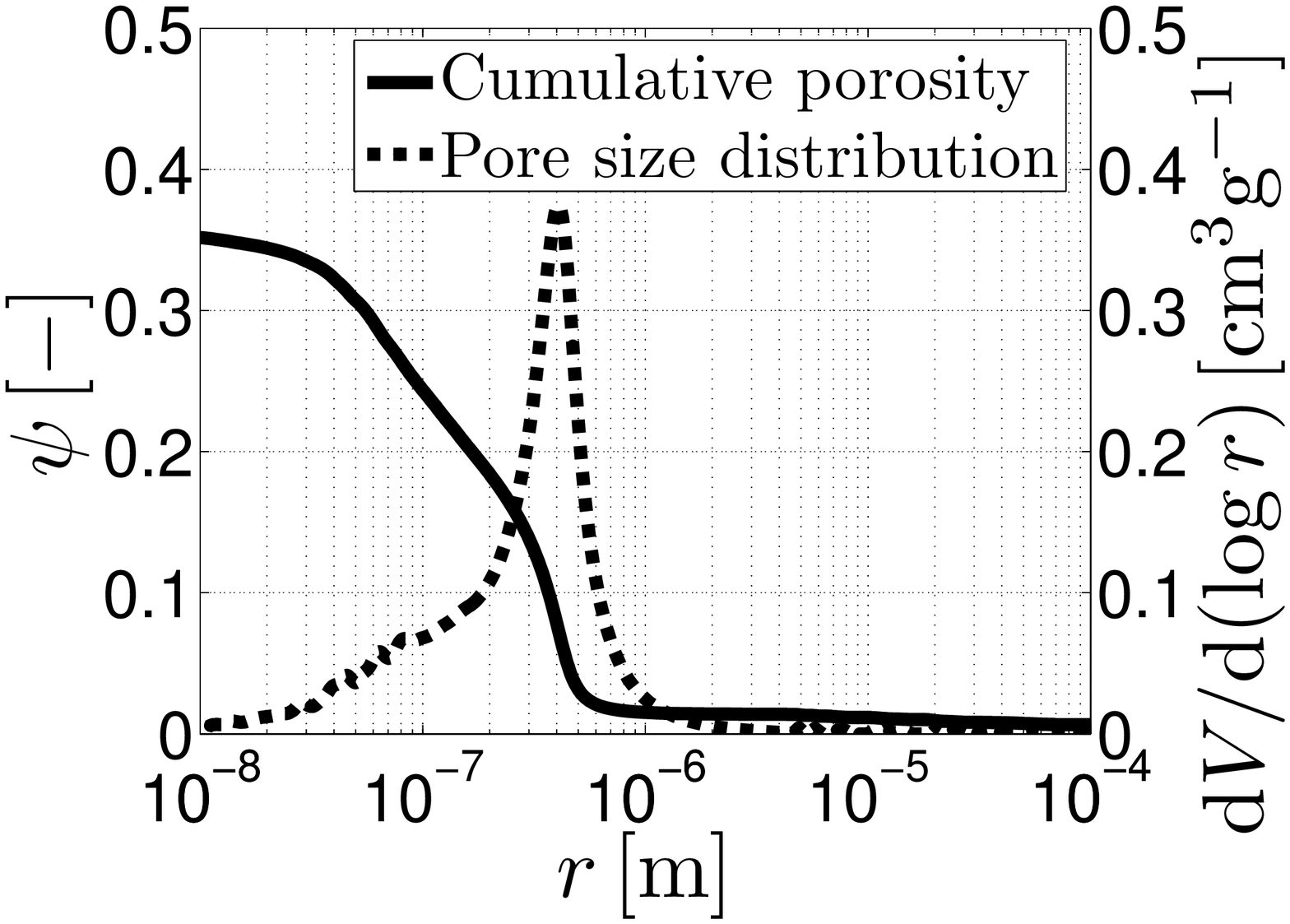}\\
(a)&(b)
\end{tabular}
\end{center}
\caption{(a) Critical pore radius $r_{\mathrm{cr}}$ as a function
of temperature, (b) pore size distribution obtained by mercury
porosimetry and resulting cumulative porosity} \label{fig:ice}
\end{figure}

The ice formation process is limited by a critical pore radius
$r_{\mathrm{cr}}(\theta)$, see Fig.~\ref{fig:ice}(a). It describes
the smallest geometrical radius of pore in which ice crystal can
form~\cite{Matala:1995,Zuber:2000:CCR},
\begin{equation}
r_{\mathrm{cr}}(\theta)=r_{\mathrm{ir}}(\theta)+r_{\mathrm{ar}}(\theta),
\end{equation}
where $r_{\mathrm{ir}}\,\mathrm{[m]}$ is the curvature radius of
ice crystal (liquid-ice interface) formed at a given temperature
and $r_{\mathrm{ar}}\,\mathrm{[m]}$ is the layer of adsorbed water
which cannot freeze during crystallization process. The empirical
formula for $r_{\mathrm{ar}}\,\mathrm{[m]}$ was proposed by
Fagerlund~\cite{Fagerlund:1973:MS} as
\begin{equation}
r_{\mathrm{ar}}(\theta)=1.97\cdot 10^{-9}
\sqrt[3]{\frac{1}{|\theta|}} \qquad \mathrm{for} \,\, \theta <
0\,\mathrm{[^{\circ}C]}
\end{equation}
and $r_{\mathrm{ir}}\,\mathrm{[m]}$ is introduced through the
simplified form of the Gibbs-Duhem equation~\cite{Wardeh:2008:CBM}
as
\begin{equation}
r_{\mathrm{ir}}(\theta)=\frac{2\gamma_{\mathrm{li}}}{\Delta
s_{\mathrm{m}}|\theta|} \qquad \mathrm{for} \,\, \theta <
0\,\mathrm{[^{\circ}C]},
\end{equation}
where $\gamma_{\mathrm{li}}\,\mathrm{[Nm^{-1}]}$ is the liquid/ice
surface tension and $\Delta s_{\mathrm{m}}\,\mathrm{[PaK^{-1}]}$
is the melting entropy.

The concept of the pore pressure caused by the ice crystals is
well known,
see~\cite{Coussy:2007:CG,Coussy:2008:CCR,Fagerlund:1973:MS,Matala:1995,Scherer:1993:JNS,Scherer:1999:CCR,Sun:2010:CCR,Wardeh:2008:CBM,Zuber:2000:CCR}.
To be more specific, let us consider spherical liquid-ice
interface at the entrance of the pore and cylindrical shape of
pores. Two interface equilibrium conditions are assumed to
describe the interaction between ice crystals and pore walls. The
Laplace relation is applicable to control the interface between
the ice crystal and the liquid water:
\begin{equation}
p_{\mathrm{i}}-p_{\mathrm{l}}=\frac{2\gamma_{\mathrm{li}}}{r_{\mathrm{ir}}(\theta)},
\label{eq:pp01}
\end{equation}
where $p_{\mathrm{l}}\,\mathrm{[Pa]}$ is the liquid pressure and
$p_{\mathrm{i}}\,\mathrm{[Pa]}$ is the pressure in the ice
crystal. The second interface relation is expressed in the form of
mechanical equilibrium between the ice crystal and the pore
pressure exerted by the ice crystal,
$p_{\mathrm{p}}\,\mathrm{[Pa]}$, as
\begin{equation}
p_{\mathrm{i}}-p_{\mathrm{p}}=\frac{\gamma_{\mathrm{li}}}{r-r_{\mathrm{ar}}(\theta)}.
\label{eq:pp02}
\end{equation}
Finally, combining Eq.~(\ref{eq:pp01}) and Eq.~(\ref{eq:pp02}), we
can write
\begin{equation}
p_{\mathrm{p}}=p_{\mathrm{l}}+ \chi(r,\theta), \label{eq:pp03}
\end{equation}
where $\chi(r,\theta)\,\mathrm{[Pa]}$ is the local pressure on the
frozen pore walls due to the ice formation and it is characterized
by following relation,
see~\cite{Multon:2012:IJNAMG,Zuber:2000:CCR},
\begin{equation}
\chi(r,\theta) =
\gamma_{\mathrm{li}}\left(\frac{2}{r_{\mathrm{ir}}(\theta)}-\frac{1}{r-r_{\mathrm{ar}}(\theta)}\right).
\end{equation}

It has been advocated in ~\cite{Scherer:1993:JNS} and
~\cite{Zuber:2000:CCR} that the average pore pressure exerted by
the ice crystal on the pore walls can be introduced as
\begin{equation}
p_{\mathrm{p}}=p_{\mathrm{l}}+\frac{1}{n}\int_{r_{\mathrm{cr}}(\theta)}^{\infty}
\chi(r,\theta)\frac{\mathrm{d}\psi}{\mathrm{d}r}\,\mathrm{d}r,
\label{eq:icepress}
\end{equation}
where $n\,\mathrm{[-]}$ is the total porosity and
$\psi(r)\,\mathrm{[-]}$ is the cumulative porosity. The cumulative
porosity $\psi\,\mathrm{[-]}$ with a pore radius greater than
$r\,\mathrm{[m]}$ is defined as, see Fig.~\ref{fig:ice}(b),
\begin{equation}
\psi(r)=\int_{r}^{\infty}\frac{\mathrm{d}\psi}{\mathrm{d}r}\,\mathrm{d}r.
\end{equation}

\subsection{Mechanical (damage) model}
The description of mechanical behavior of a porous media saturated
with a liquid water was firstly proposed by
Biot~\cite{Biot:2002:JAP} and extended in the more general context
of continuum thermodynamics for the ice crystals by
Coussy~\cite{Coussy:2007:CG,Coussy:2008:CCR}. According to this
approach, the formula between the effective stress
$\vek{\sigma}'\,\mathrm{[Pa]}$ and total stress
$\vek{\sigma}\,\mathrm{[Pa]}$ has following form
\begin{equation}
\vek{\sigma} = \vek{\sigma}'-bp_{\mathrm{p}}\vek{i},
\label{eq:dm01}
\end{equation}
where $b\,\mathrm{[-]}$ is Biot's coefficient,
$p_{\mathrm{p}}\,\mathrm{[Pa]}$ is the pressure exerted by ice
crystal on the pore walls and $\vek{i}=\{1,1,1,0,0,0\}^{\sf T}$.
The standard formulation of Biot's coefficient related to the
liquid water is based on introduction of the bulk modulus of
porous material $K\,\mathrm{[Pa]}$ and the bulk modulus of solid
matrix $K_{\mathrm{s}}\,\mathrm{[Pa]}$. For our purpose, we
utilize more convenient relation for porous system derived in
~\cite{Sun:2010:CCR} .
\begin{equation}
b = \frac{2n}{n+1}\approx 1 - \frac{K}{K_{\mathrm{s}}},
\end{equation}
where $n\,\mathrm{[-]}$ is the total porosity. As a further
extension, we introduce one parameter isotropic nonlocal damage
model given by constitutive law,
see~\cite{Bazant:2002:JEM,Horak:2009},
\begin{equation}
\vek{\sigma}' = (1-d_{w})\mat{D}_{e}\vek{\varepsilon},
\label{eq:dm02}
\end{equation}
damage law
\begin{equation}
d_{w}=g(\kappa)
\end{equation}
and loading-unloading conditions
\begin{equation}
f(\vek{\varepsilon},\kappa)=\varepsilon_{\mathrm{eq}}(\vek{\varepsilon})-\kappa
\leq 0, \qquad \dot{\kappa} \geq 0, \qquad
f(\vek{\varepsilon},\kappa)\dot{\kappa} = 0,
\end{equation}
where $\mat{D}_{e}\,\mathrm{[Pa]}$ is the elastic stiffness
matrix, $d_{w}\,\mathrm{[-]}$ is the damage parameter,
$\kappa\,\mathrm{[-]}$ is the internal variable corresponding to
the maximum value of equivalent strain
$\varepsilon_{\mathrm{eq}}\,\mathrm{[-]}$ reached in the loading
history. The equivalent strain can be expressed in Mazars's form
as
\begin{equation}
\varepsilon_{\mathrm{eq}}=\sqrt{\sum_{I=1}^{3}\langle
\varepsilon_{I}\rangle ^{2}},
\end{equation}
where $\varepsilon_{I}\,\mathrm{[-]}$ is component of the
principal strains and the brackets denote the positive part.
According to ~\cite{Bazant:2002:JEM}, the local value
$\varepsilon_{\mathrm{eq}}\,\mathrm{[-]}$ is replaced by its
nonlocal average defined as
\begin{equation}
\varepsilon_{\mathrm{eq,nonlocal}}(\vek{x})
=\int_{V}\phi(\vek{x},\vek{\xi})\varepsilon_{\mathrm{eq}}(\vek{\xi})\mathrm{d}\vek{\xi},
\end{equation}
where $\phi(\vek{x},\vek{\xi})\,\mathrm{[-]}$ is the nonlocal
weight function representing distance in the domain between the
source point $\vek{\xi}\,\mathrm{[m]}$ and target point
$\vek{x}\,\mathrm{[m]}$. For more details about nonlocal
formulation we refer to~\cite{Bazant:2002:JEM}. Finally, the
damage law $d_{w}=g(\kappa)$ is provided by following relation
\begin{equation}
g(\kappa)=\left\{\begin{array}{ll} 0 & \mathrm{for}\,\, 0\leq\kappa\leq\varepsilon_{0}, \\
1 - \frac{\kappa-\varepsilon_{0}}{\varepsilon_{f}-\varepsilon_{0}} & \mathrm{for}\,\, \varepsilon_{0}\leq\kappa\leq\varepsilon_{f}, \\
1 & \mathrm{for}\,\, \varepsilon_{f}\leq\kappa.
\end{array}\right.,
\end{equation}
where $\varepsilon_{f}\,\mathrm{[-]}$ is the equivalent strain at
critical crack opening and $\varepsilon_{0}\,\mathrm{[-]}$ is the
strain at the elastic limit.

Considering the stress relation Eq.~(\ref{eq:dm01}) and
Eq.~(\ref{eq:dm02}), the linear momentum balance equation for the
porous system, can be expressed in the following form:
\begin{equation}
\vek{\nabla}\cdot[\vek{\sigma}'-bp_{\mathrm{p}}\vek{i}]+\vek{b}=0,
\label{eq:linmom}
\end{equation}
where $\vek{b}\,\mathrm{[Nm^{-3}]}$ is the body force. The
governing equations are discretized in space using the standard
finite element approximation. The unknown displacement field
$\vek{u}\,\mathrm{[m]}$ is expressed in terms of its nodal values.
Details of the formulation may be found
in~\cite{Gawin:2003:CMAME}.

\subsection{Boundary and initial conditions}
{\color{red} To complete the proposed material model, the initial
and boundary conditions are set as follows:}
\begin{itemize}
\item The Dirichlet boundary conditions
\begin{eqnarray}
\theta&=&\hat{\theta}(t)\quad\mathrm{on}\quad\Gamma_{\theta}^{\mathrm{I}},\nonumber\\
\varphi&=&\hat{\varphi}(t)\quad\mathrm{on}\quad\Gamma_{\varphi}^{\mathrm{I}},\\
\vek{u}&=&\hat{\vek{u}}(t)\quad\mathrm{on}\quad\Gamma_{u}^{\mathrm{I}},\nonumber
\end{eqnarray}
\item The Neumann boundary conditions
\begin{eqnarray}
{[-\lambda\vek{\nabla}\theta-\delta_{\mathrm{v}}\vek{\nabla}\{\varphi
p_{\mathrm{sat}}(\theta)\}]}\cdot\vek{n}&=&
q_{\mathrm{h}}(t)+q_{\mathrm{h,swr}}(t)
\quad\mathrm{on}\quad\Gamma_{\theta}^{\mathrm{II}},\nonumber \\
{[-D_{\varphi}\vek{\nabla}\varphi-\delta_{\mathrm{v}}\vek{\nabla}\{
\varphi p_{\mathrm{sat}}(\theta)\}]}\cdot \vek{n}&=& q_{\mathrm{v}}(t)+q_{\mathrm{l}}(t)+q_{\mathrm{l,rain}}(t)\quad\mathrm{on}\quad\Gamma_{\varphi}^{\mathrm{II}},\nonumber\\
\vek{\sigma}'\cdot
\vek{n}&=&\vek{\sigma}_{\mathrm{t}}(t)\quad\mathrm{on}\quad\Gamma_{u}^{\mathrm{II}},
\end{eqnarray}
\item The Robin boundary conditions
\begin{eqnarray}
{[-\lambda\vek{\nabla}\theta-\delta_{\mathrm{v}}\vek{\nabla}\{\varphi
p_{\mathrm{sat}}(\theta)\}]}\cdot\vek{n}&=&\alpha_{\mathrm{h}}{[\theta-\theta_{\infty}(t)]}
\quad\mathrm{on}\quad\Gamma_{\theta}^{\mathrm{III}},\nonumber \\
{[-D_{\varphi}\vek{\nabla}\varphi-\delta_{\mathrm{v}}\vek{\nabla}\{
\varphi p_{\mathrm{sat}}(\theta)\}]}\cdot \vek{n}&=&
\beta_{\mathrm{v}}{[\varphi-\varphi_{\infty}(t)]}\quad\mathrm{on}\quad\Gamma_{\varphi}^{\mathrm{III}},
\end{eqnarray}
\item Initial conditions
\begin{equation}
\theta(\vek{x},0)=\theta_{\mathrm{in}}\quad
\varphi(\vek{x},0)=\varphi_{\mathrm{in}}\quad
\vek{u}(\vek{x},0)=\vek{u}_{\mathrm{in}}\quad\mathrm{for}
\,\,\mathrm{all}\quad\vek{x}\in\Omega,
\end{equation}
\end{itemize}
where the symbol $\,\hat{\cdot}\,$ denotes prescribed value,
$\vek{n}\,\mathrm{[-]}$ is the unit normal vector,
$q_{\mathrm{h,swr}}\,\mathrm{[Wm^{-2}]}$ is the solar short-wave
radiation flux, $q_{\mathrm{l,rain}}\,\mathrm{[kgm^{-2}s^{-1}]}$
is the driving-rain flux,
$\vek{\sigma}_{\mathrm{t}}\,\mathrm{[Pa]}$ is the imposed
traction, $\alpha_{\mathrm{h}}\,\mathrm{[Wm^{-2}K^{-1}]}$ is the
heat transfer coefficient,
$\beta_{\mathrm{v}}\,\mathrm{[kgm^{-2}s^{-1}Pa^{-1}]}$ is the
water vapor transfer coefficient, $\theta_{\infty}$ and
$\varphi_{\infty}$ are the ambient temperature and moisture,
respectively.
\begin{figure} [h!]
\begin{center}
\begin{tabular}{c}
\includegraphics*[width=80mm,keepaspectratio]{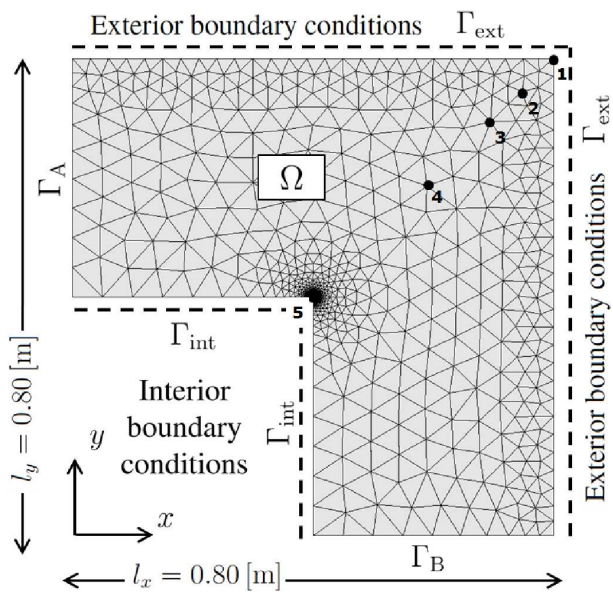}\\
\end{tabular}
\end{center}
\caption{$2$-D domain with initial and boundary conditions}
\label{fig:schema}
\end{figure}

\section{Numerical example}
\label{sec:ne}

In this section, we employ the proposed thermo-hygro-mechanical
model to perform a numerical simulation of transport processes
below freezing point in porous media and their impact on the
mechanical properties. In doing so we consider geometry together
with the initial and loading conditions displayed in
Fig~\ref{fig:schema}. Two-dimensional L-shape domain was
discretized by an FE mesh into $741$ nodes and $1358$ triangular
elements. The solution of the time-dependent problem also involves
a discretization of the time domain into $744$ uniform time steps
chosen with regard to the convergence criteria of nonlinear
solution.

\renewcommand{\arraystretch}{1.1}
\begin{table}[h!]
\begin{center}
\begin{tabular}{lllcl}
 Symbol & Unit & & Value & Ref. \\
\hline \multicolumn{5}{l}{\textit{Transport properties of mortar}} \\
$w_{f}$ &  $\mathrm{[kgm^{-3}]}$ &free water saturation  & 160 & ~\cite{Sykora:2012:JCAM} \\
$w_{\mathrm{80}}$  & $\mathrm{[kgm^{-3}]}$ & water content at $\varphi=0.8$ [-]  & 23 & ~\cite{Sykora:2012:JCAM} \\
$\lambda_{\mathrm{0}}$ & $\mathrm{[Wm^{-1}K^{-1}]}$ & thermal conductivity  & 0.45 & ~\cite{Sykora:2012:JCAM} \\
$b_{\mathrm{tcs}}$  & $\mathrm{[-]}$ & thermal conductivity supplement  & 9 & ~\cite{Sykora:2012:JCAM} \\
$\rho_{\mathrm{s}}$ & $\mathrm{[kgm^{-3}]}$ &  bulk density  & 1670 & ~\cite{Sykora:2012:JCAM}\\
$\mu$   & $\mathrm{[-]}$ &water vapor diffusion resistance factor  & 9.63 & ~\cite{Nunes:2012:EM,Sykora:2012:JCAM}\\
$a$ & $\mathrm{[kgm^{-2}s^{-0.5}]}$ &  water absorption coefficient  & 0.82 & ~\cite{Nunes:2012:EM,Sykora:2012:JCAM}\\
$c_{\mathrm{s}}$ & $\mathrm{[Jkg^{-1}K^{-1}]}$ &specific heat capacity  & 1000 & ~\cite{Nunes:2012:EM,Sykora:2012:JCAM}\\
\hline \multicolumn{5}{l}{\textit{Mechanical properties of mortar}} \\
$E$ & $\mathrm{[Pa]}$ & Young's modulus  & $1\cdot 10^{10}$ & ~\cite{Milani:2012:CBM,Milani:2012:IJSS}\\
$\nu$ & $\mathrm{[-]}$ & Poisson's ratio  & 0.2 & ~\cite{Milani:2012:CBM,Milani:2012:IJSS}\\
$f_{t}$ & $\mathrm{[Pa]}$ & tensile strength  & $2.5\cdot 10^{6}$ & ~\cite{Milani:2012:CBM,Milani:2012:IJSS}\\
$\varepsilon_f$ & $\mathrm{[-]}$ & equivalent strain at critical crack opening  & $2.5\cdot 10^{-3}$ & ~\cite{Bazant:2002:JEM,Horak:2009}\\
$l_{\mathrm{intl}}$ & $\mathrm{[m]}$ & internal length  & $1\cdot 10^{-3}$ & ~\cite{Bazant:2002:JEM,Horak:2009}\\
$\alpha$ & $\mathrm{[K^{-1}]}$ & thermal expansion coefficient  & $1.2\cdot 10^{-5}$ & ~\cite{Milani:2012:CBM,Milani:2012:IJSS}\\
\hline \multicolumn{5}{l}{\textit{Ice formation process}} \\
$\gamma_{\mathrm{li}}$ & $\mathrm{[Nm^{-1}]}$ & liquid/ice surface tension & 0.0409 & ~\cite{Liu:2011:CRT}\\
$\Delta s_{\mathrm{m}}$ & $\mathrm{[PaK^{-1}]}$ & melting entropy & $1.2\cdot 10^{6}$ & ~\cite{Liu:2011:CRT} \\
$n$ & $\mathrm{[-]}$ & total porosity & 0.35 & ~\cite{Nunes:2012:EM,Slizkova:2010:ROUD} \\
$\psi$ & $\mathrm{[-]}$ & cumulative volume of pores & Fig.~\ref{fig:ice}(b) & ~\cite{Nunes:2012:EM,Slizkova:2010:ROUD} \\
\hline \multicolumn{5}{l}{\textit{Other properties}} \\
$\alpha_{\mathrm{h}}$ & $\mathrm{[Wm^{-2}K^{-1}]}$ & heat transfer coefficient & 8 & ~\cite{Kunzel:1996:IJHMT}\\
$\beta_{\mathrm{v}}$ & $\mathrm{[kgm^{-2}s^{-1}Pa^{-1}]}$ & water vapor transfer coefficient & $5.6\cdot 10^{-8}$ & ~\cite{Kunzel:1996:IJHMT}\\
$\alpha_{\mathrm{swr}}$ & $\mathrm{[-]}$ & short-wave absorption coefficient & 0.6 & ~\cite{Kunzel:1996:IJHMT}\\
\hline
\end{tabular}
\caption{Input parameters of numerical simulation}
\label{tab:matpar}
\end{center}
\end{table}

The measured material parameters, corresponding to a lime mortar,
are listed in Tab.~\ref{tab:matpar}. Several parameters were
obtained from a set of experimental measurements providing mostly
the hygric and thermal properties of mortar, see
~\cite{Nunes:2012:EM}. Unfortunately, no additional experiments
were conducted for the mechanical properties, hence we utilize
some material parameters of mortars mentioned in the literature,
see Tab.~\ref{tab:matpar}.

\begin{figure} [h!]
\begin{center}
\begin{tabular}{cc}
\includegraphics*[width=75mm,keepaspectratio]{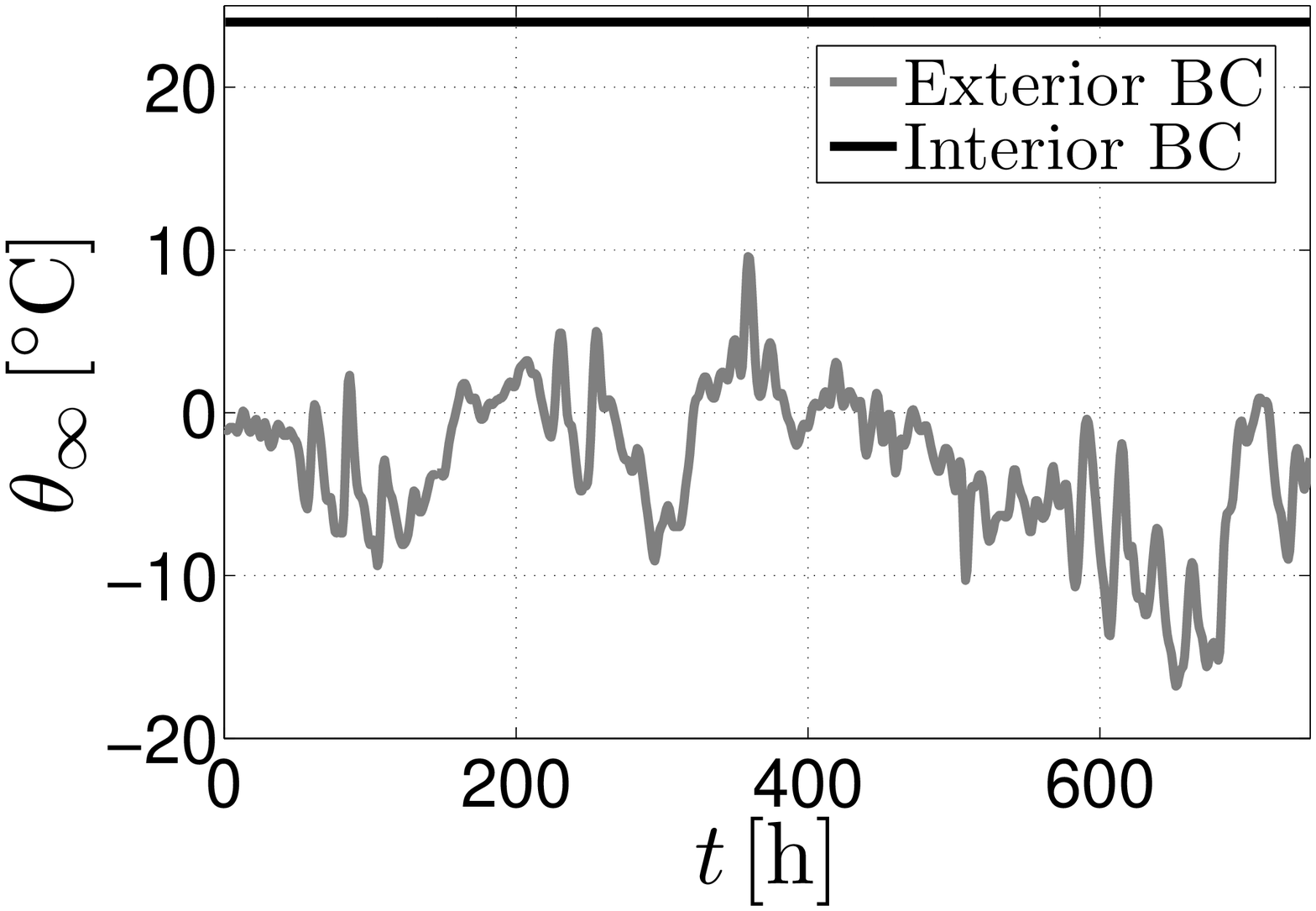}&
\includegraphics*[width=75mm,keepaspectratio]{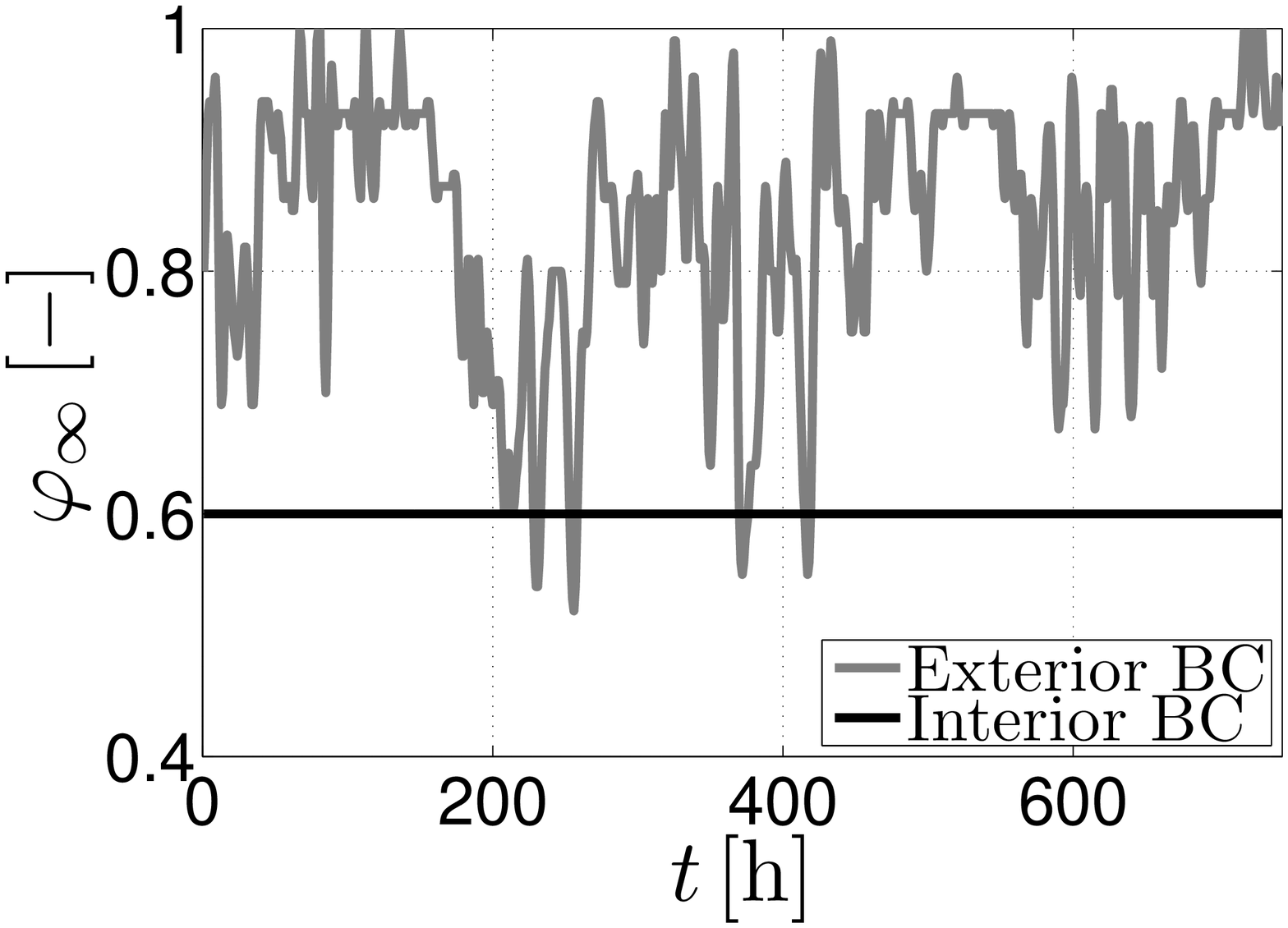}\\
(a)&(b) \\
\includegraphics*[width=75mm,keepaspectratio]{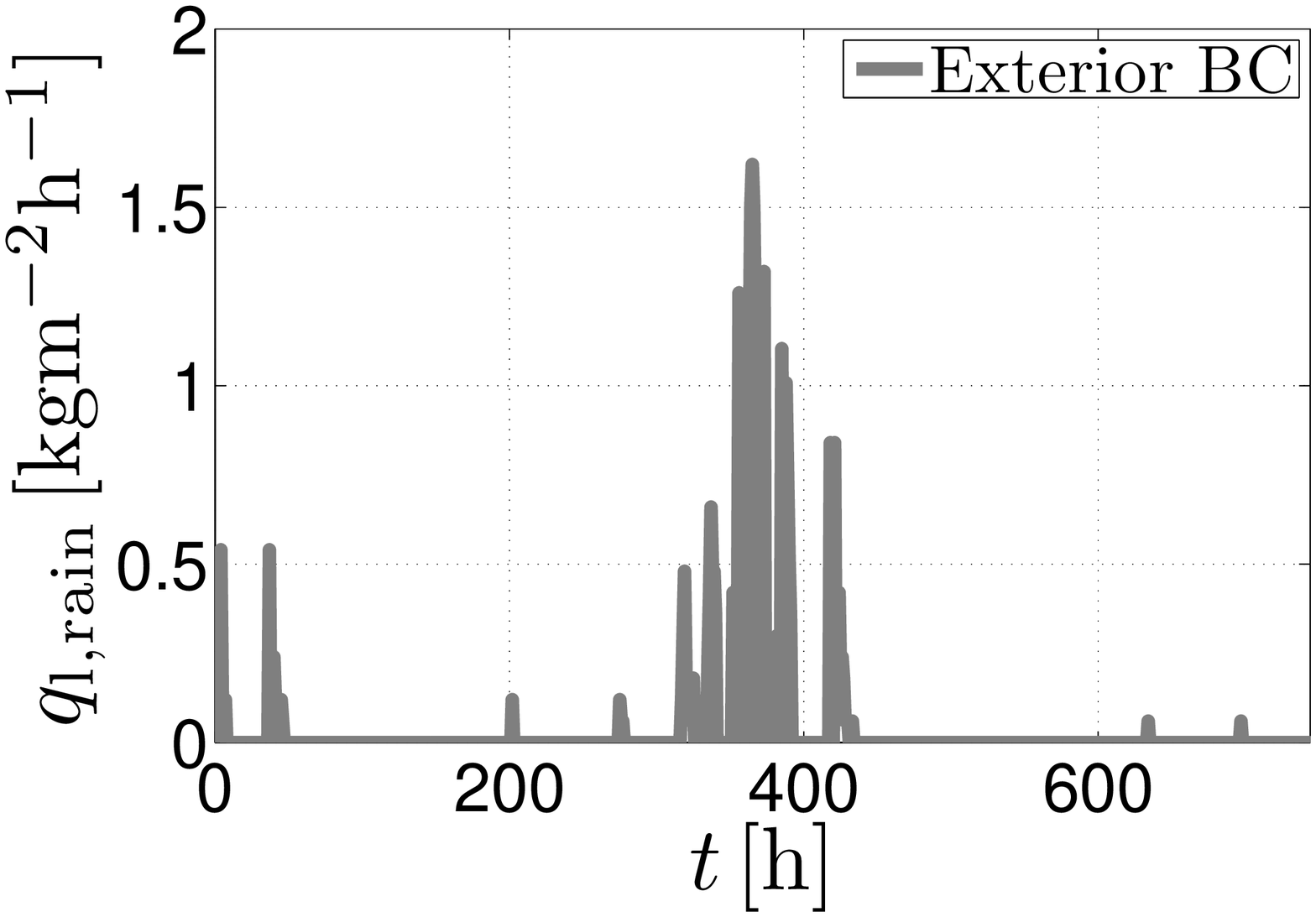}&
\includegraphics*[width=75mm,keepaspectratio]{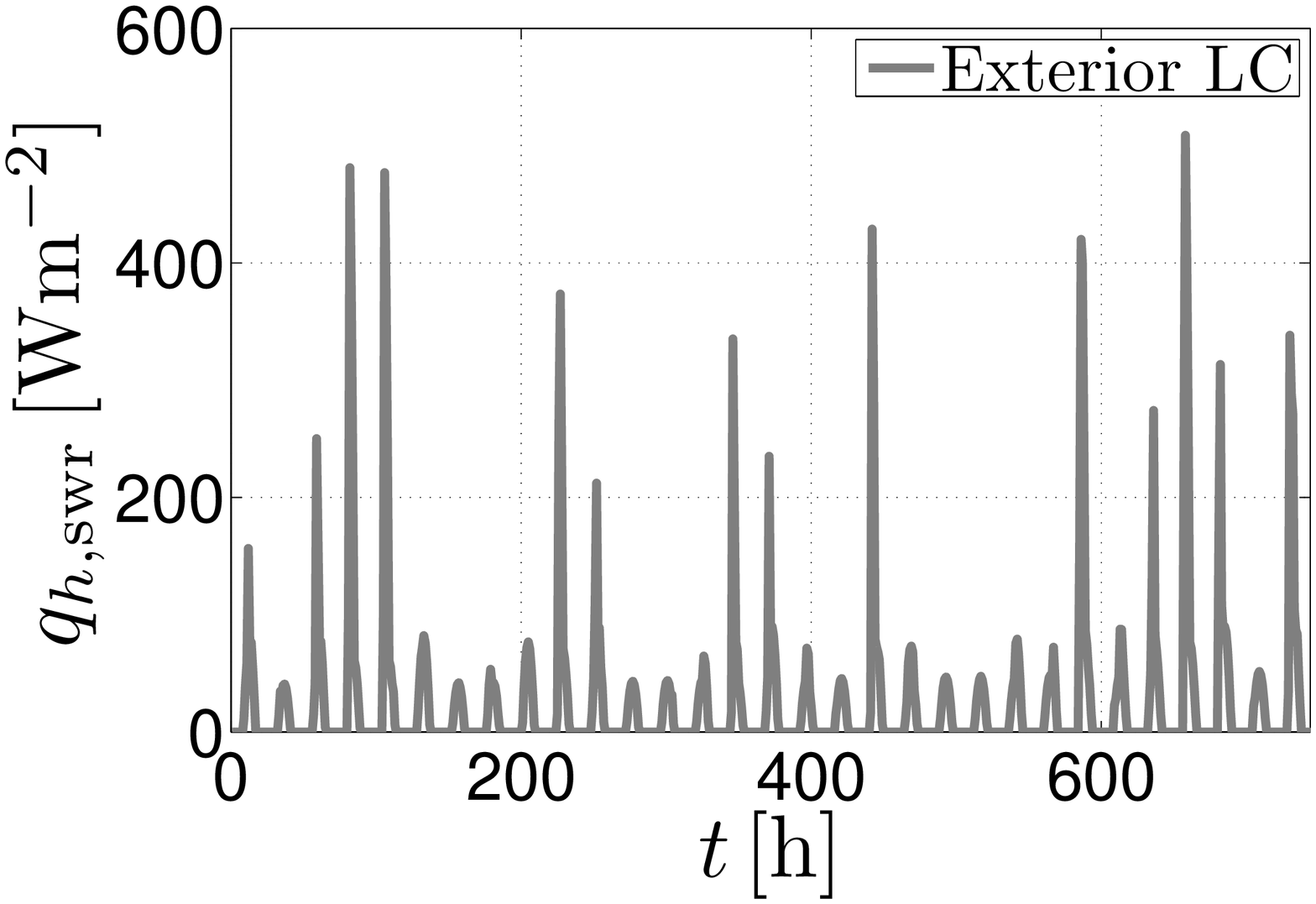}\\
(c)&(d) \\
\end{tabular}
\end{center}
\caption{Exterior and interior boundary conditions - (a)
temperature, (b) moisture, (c) driving-rain, (d) short-wave
radiation} \label{fig:bc}
\end{figure}

The initial conditions were set equal to
$\vek{u}_{\mathrm{in}}=\vek{0}\,\mathrm{[m]}$,
$\theta_{\mathrm{in}}=14\,\mathrm{[^{\circ} C]}$ and
$\varphi_{\mathrm{in}}=0.5\,\mathrm{[-]}$ in the whole domain. The
following Robin boundary conditions were imposed: on the interior
side a constant temperature of $24\,\mathrm{[^{\circ}C]}$ and a
constant relative humidity $0.6\,\mathrm{[-]}$ were maintained,
while on the exterior side the real climatic data representing the
winter conditions were prescribed, see Figs.~\ref{fig:bc}(a),(b).
Moreover, the exterior side of the domain was loaded by the heat
flux from solar short-wave radiation
$q_{\mathrm{h,swr}}\,\mathrm{[Wm^{-2}]}$ and the driving-rain flux
$q_{\mathrm{l,rain}}\,\mathrm{[kgm^{-2}s^{-1}]}$ (the Neumann
boundary conditions), see Figs.~\ref{fig:bc}(c),(d) and
Tab.~\ref{tab:BC}.
\begin{table}[h!]
\begin{center}
\begin{tabular}{lllcl}
Side & BC type & Description & Fig.\\
\hline \multicolumn{4}{l}{\textit{Boundary conditions}} \\
$\Gamma_{\mathrm{ext}}$ & II &
${[-\lambda\vek{\nabla}\theta-\delta_{\mathrm{v}}\vek{\nabla}\{\varphi
p_{\mathrm{sat}}(\theta)\}]}\cdot\vek{n} =q_{\mathrm{h,swr}}(t)$ &
\ref{fig:bc}(d)
\\
$\Gamma_{\mathrm{ext}}$ & II &
${[-D_{\varphi}\vek{\nabla}\varphi-\delta_{\mathrm{v}}\vek{\nabla}\{
\varphi p_{\mathrm{sat}}(\theta)\}]}\cdot
\vek{n}=q_{\mathrm{l,rain}}(t)$ & \ref{fig:bc}(c)
\\
$\Gamma_{\mathrm{ext}}$ & III &
${[-\lambda\vek{\nabla}\theta-\delta_{\mathrm{v}}\vek{\nabla}\{\varphi
p_{\mathrm{sat}}(\theta)\}]}\cdot\vek{n}=8{[\theta-\theta_{\infty}(t)]}$
& \ref{fig:bc}(a)
\\
$\Gamma_{\mathrm{ext}}$ & III &
${[-D_{\varphi}\vek{\nabla}\varphi-\delta_{\mathrm{v}}\vek{\nabla}\{
\varphi p_{\mathrm{sat}}(\theta)\}]}\cdot \vek{n}= 5.6\cdot
10^{-8}{[\varphi-\varphi_{\infty}(t)]}$ & \ref{fig:bc}(b)
\\
$\Gamma_{\mathrm{int}}$ & III &
${[-\lambda\vek{\nabla}\theta-\delta_{\mathrm{v}}\vek{\nabla}\{\varphi
p_{\mathrm{sat}}(\theta)\}]}\cdot\vek{n}=8{[\theta-24]}$ &
\ref{fig:bc}(a)
\\
$\Gamma_{\mathrm{int}}$ & III &
${[-D_{\varphi}\vek{\nabla}\varphi-\delta_{\mathrm{v}}\vek{\nabla}\{
\varphi p_{\mathrm{sat}}(\theta)\}]}\cdot \vek{n}= 5.6\cdot
10^{-8}{[\varphi-0.6]}$ & \ref{fig:bc}(b)
\\
$\Gamma_{\mathrm{A}}$ & I & $u_{x}(t)=0\,\mathrm{[m]}$ &
\\
$\Gamma_{\mathrm{B}}$ & I & $u_{y}(t)=0\,\mathrm{[m]}$ &
\\
\hline \multicolumn{4}{l}{\textit{Initial conditions}} \\
$\Omega$ & & $\theta_{\mathrm{in}}=14\,\mathrm{[^{\circ} C]}$ & \\
$\Omega$ & & $\varphi_{\mathrm{in}}=0.5\,\mathrm{[-]}$ & \\
$\Omega$ & & $\vek{u}_{\mathrm{in}}=\vek{0}\,\mathrm{[m]}$ & \\
\hline
\end{tabular}
\caption{Boundary and initial conditions} \label{tab:BC}
\end{center}
\end{table}

\begin{figure} [h!]
\begin{center}
\begin{tabular}{cc}
\includegraphics*[width=75mm,keepaspectratio]{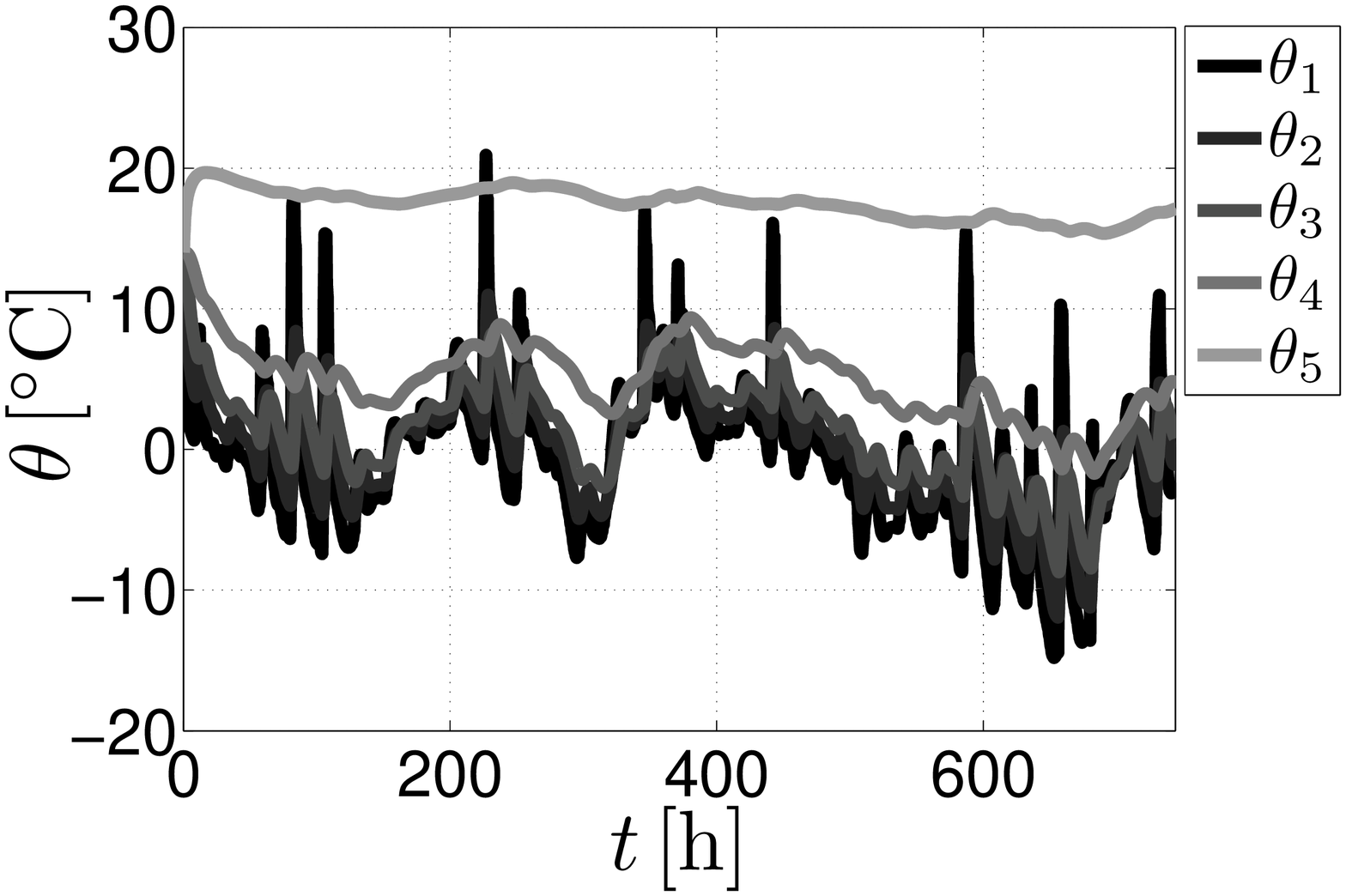}&
\includegraphics*[width=75mm,keepaspectratio]{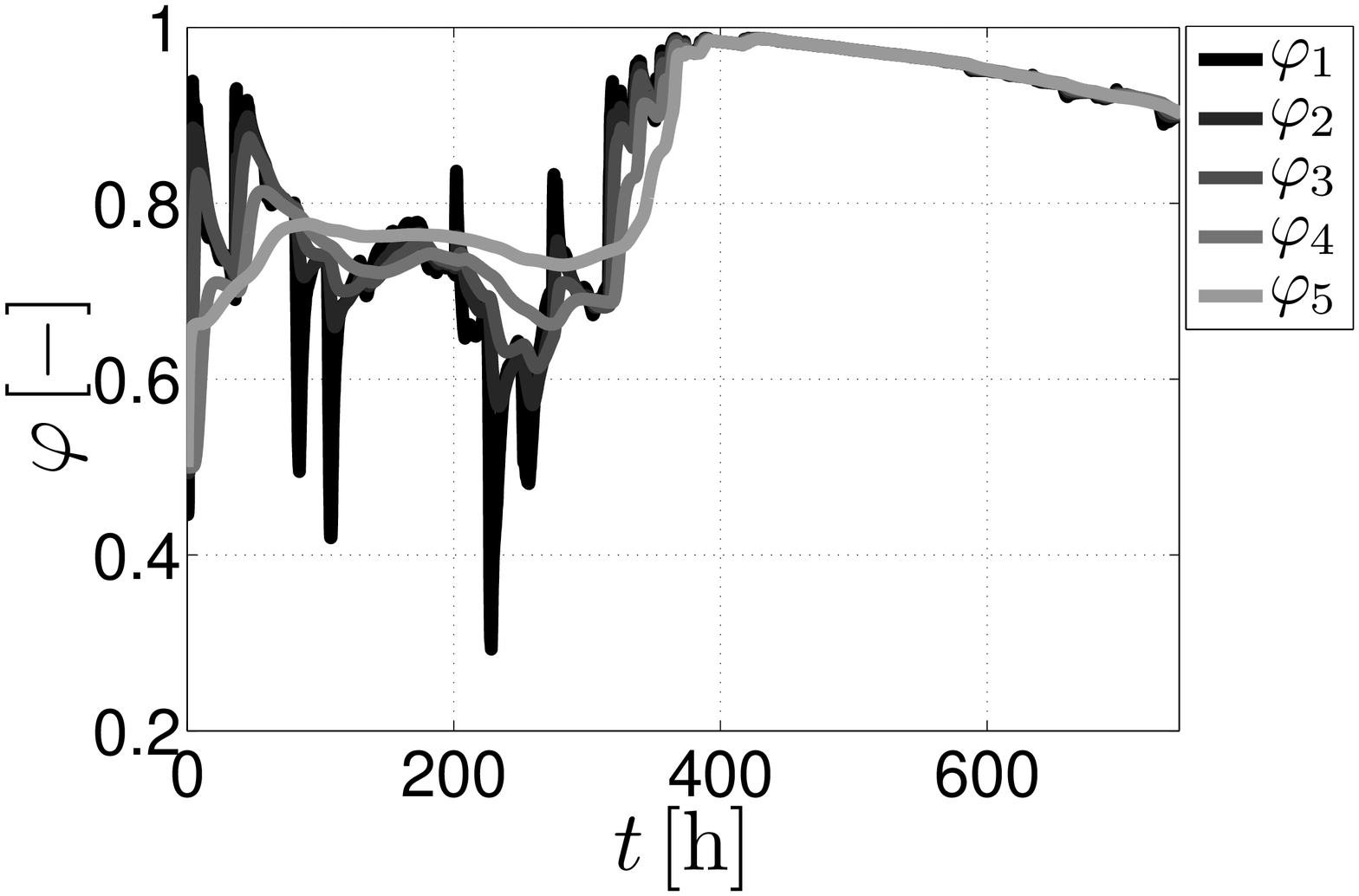}\\
(a)&(b) \\
\end{tabular}
\end{center}
\caption{(a) Resulting temperature at selected nodes, (b)
resulting moisture at selected nodes} \label{fig:res}
\end{figure}
The results are presented in Fig.~\ref{fig:res} showing variation
of the temperature and moisture at selected nodes labeled in
Fig.~\ref{fig:schema}. The obtained results clearly manifesting
the influence of exterior boundary conditions on the temperature
and moisture fields, especially near the exterior surface of the
$2$-D domain.

\begin{figure} [h!]
\begin{center}
\begin{tabular}{cc}
\includegraphics*[width=75mm,keepaspectratio]{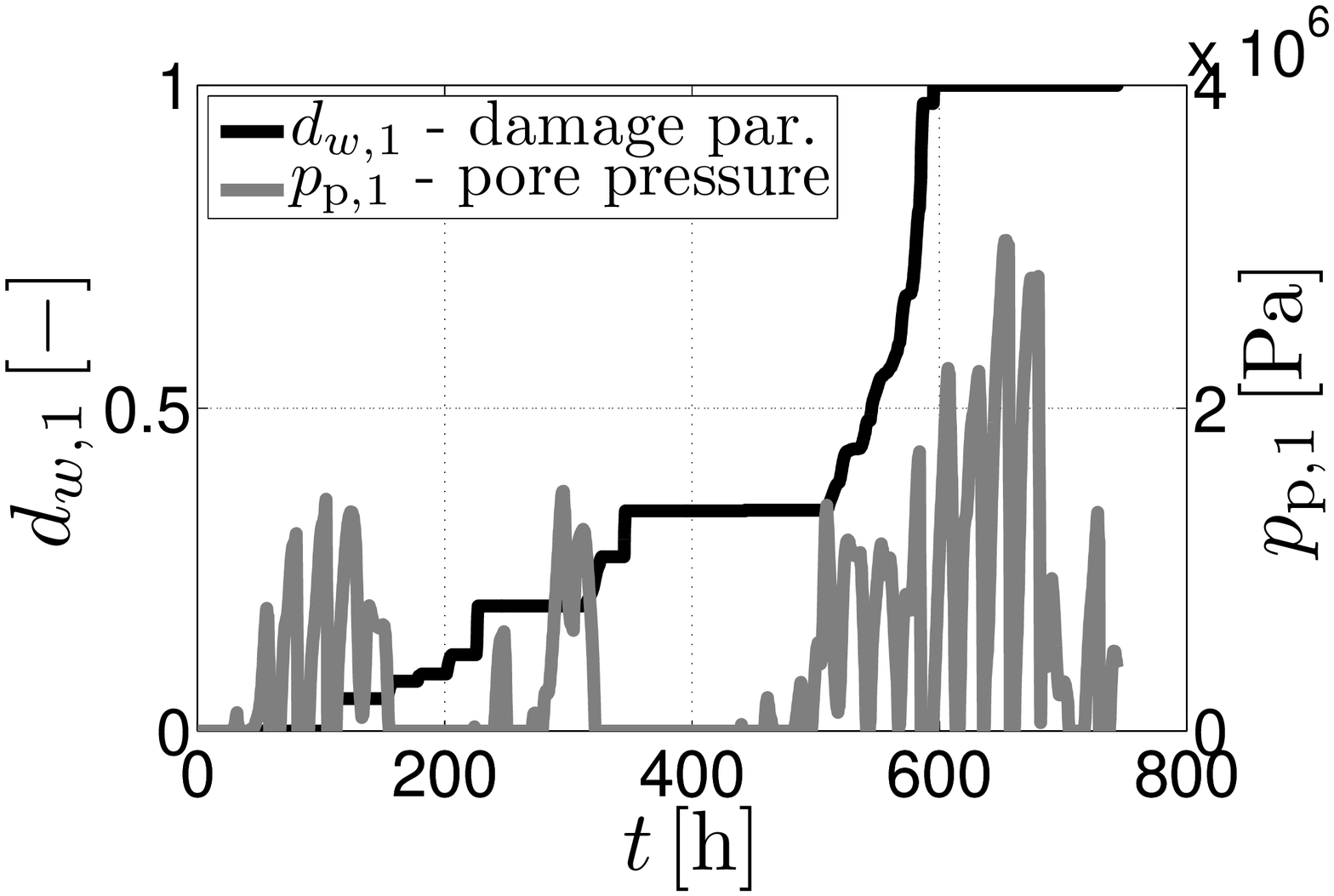}&
\includegraphics*[width=73mm,keepaspectratio]{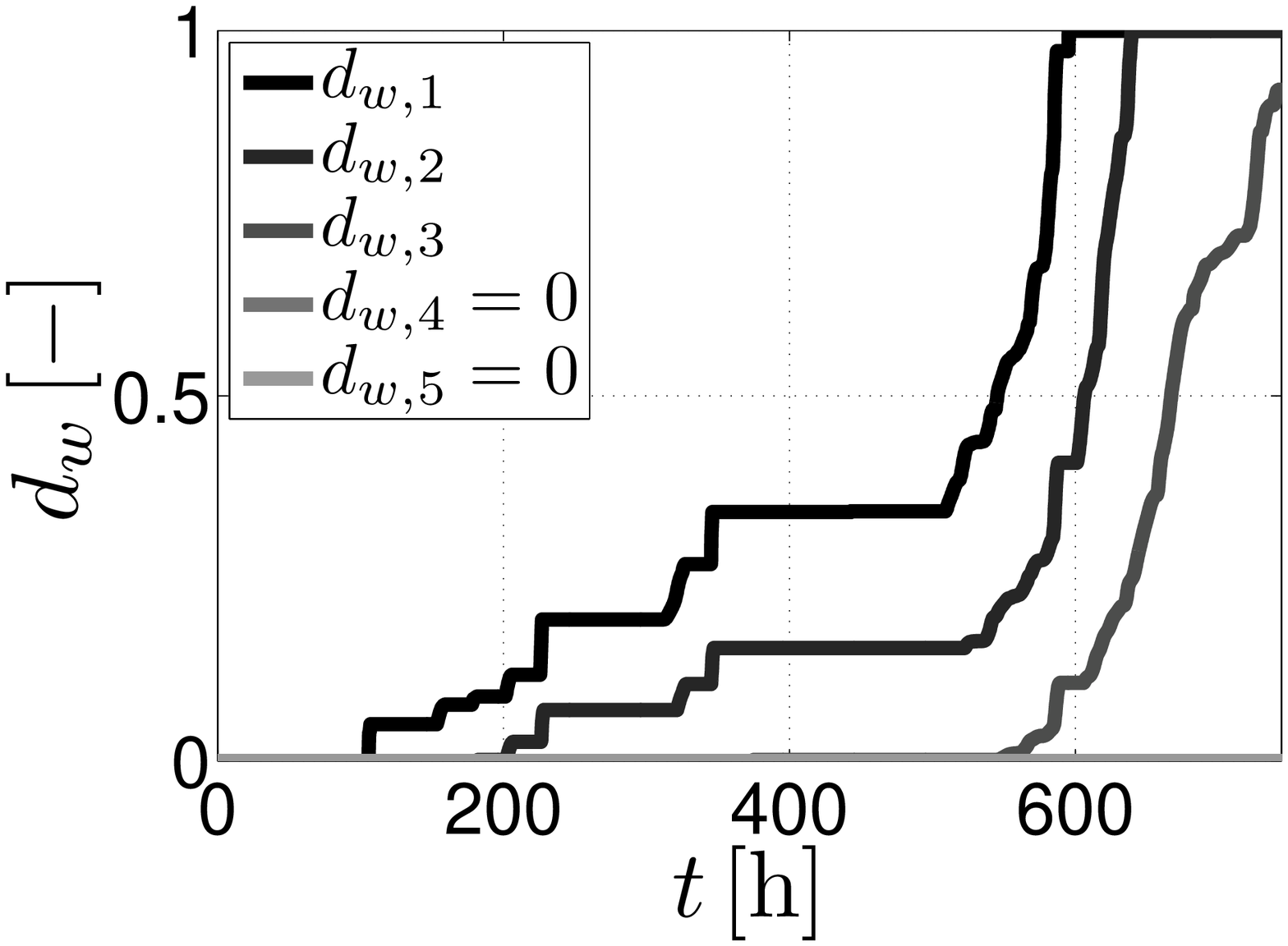}\\
(a)&(b) \\
\end{tabular}
\end{center}
\caption{(a) Evolution of damage parameter $d_{w,1}\,\mathrm{[-]}$
and pore pressure $p_{\mathrm{p},1}\,\mathrm{[Pa]}$ at node $1$,
(b) Evolution of damage parameters $d_{w}\,\mathrm{[-]}$ at
selected nodes} \label{fig:damage}
\end{figure}
\begin{figure} [h!]
\begin{center}
\begin{tabular}{cc}
\includegraphics*[width=70mm,keepaspectratio]{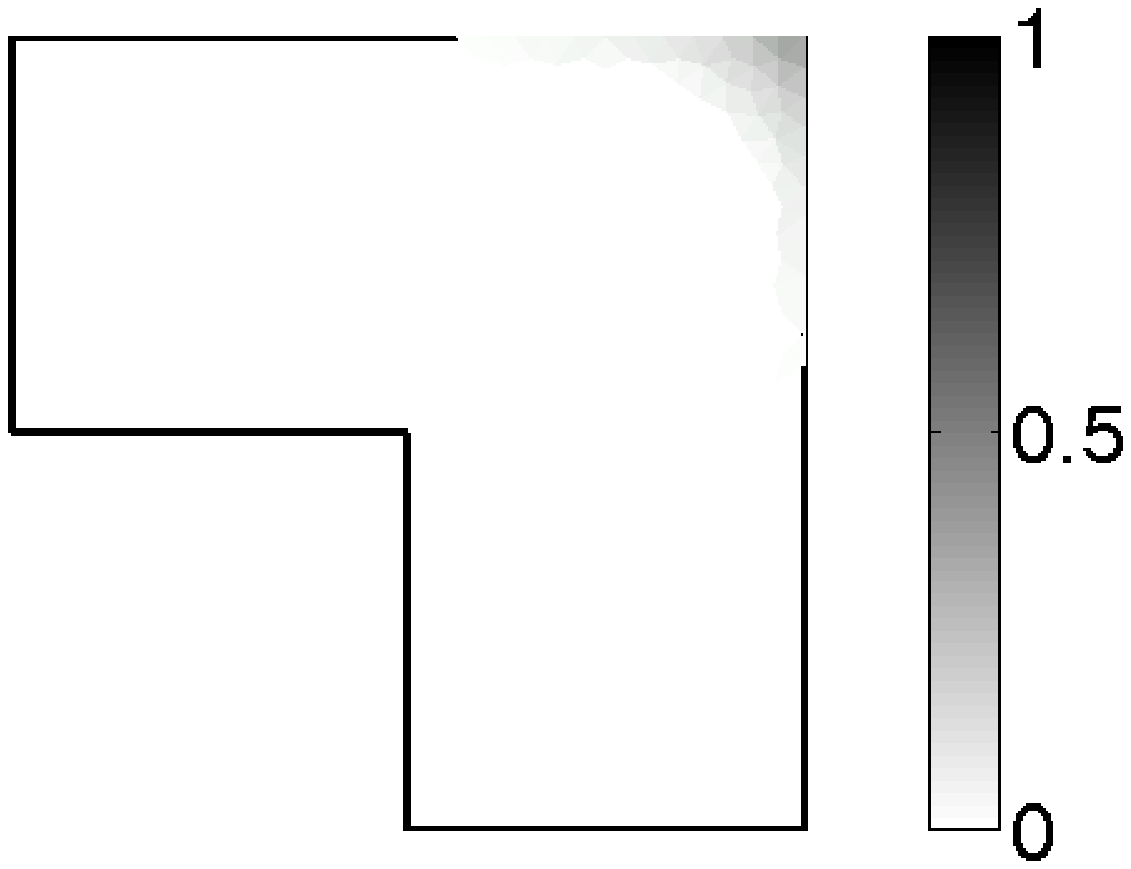}&
\includegraphics*[width=70mm,keepaspectratio]{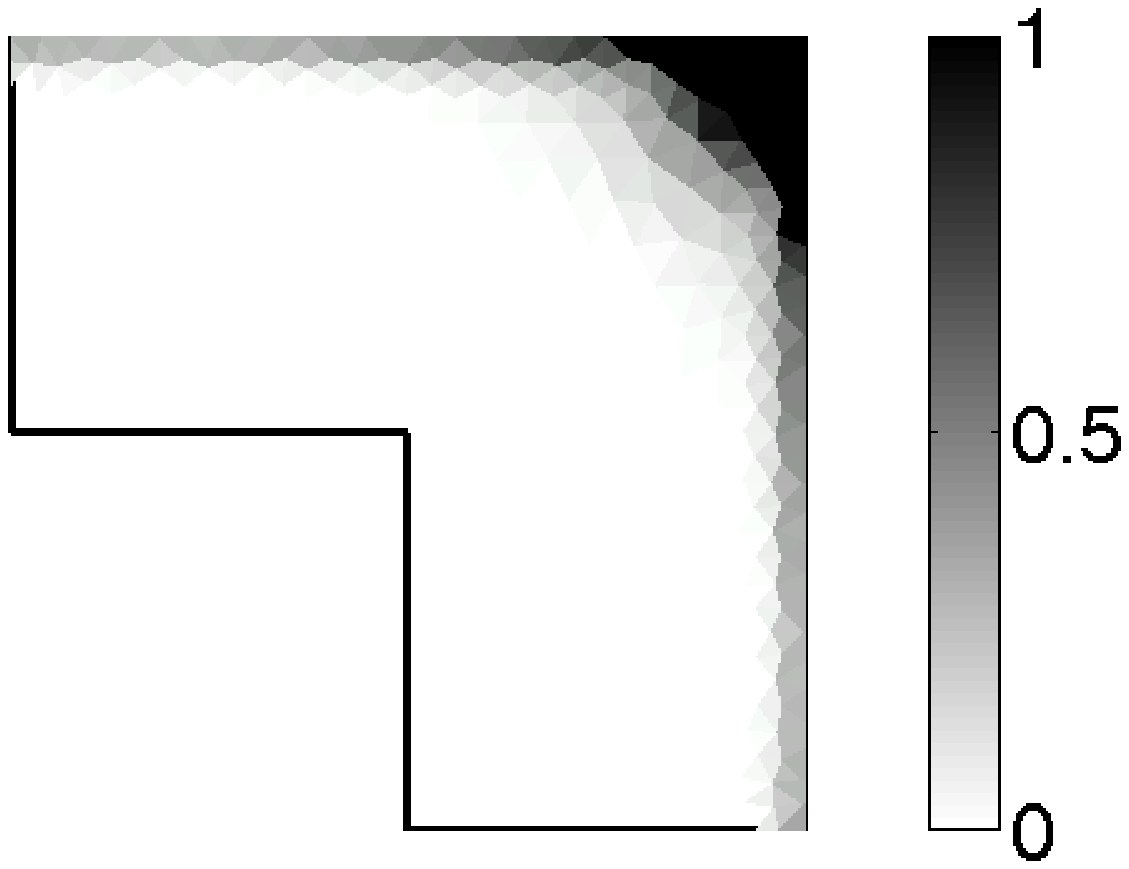}\\
(a)&(b) \\
\end{tabular}
\end{center}
\caption{(a) Evolution of damage parameter $d_{w}\,\mathrm{[-]}$
after $t=372\,\mathrm{[h]}$, (b) evolution of damage parameter
$d_{w}\,\mathrm{[-]}$ at the end of analyzed time period
($t=744\,\mathrm{[h]}$)} \label{fig:damage_time}
\end{figure}

Several interesting results have been derived within the scope of
the calculation of internal damage. Figs.~\ref{fig:damage}(a),(b)
display the evolution of damage parameter and its dependence on
the average pore pressure. Beside the comparison of the evolution
of damage parameter in the time, we also compare growth of damage
parameter in the domain, see Fig.~\ref{fig:damage_time}. Analysis
of these results allows better understanding of physical phenomena
in porous media subjected to the frost action. A fast moisture
increase in the zone close to the exterior surface
(Fig.~\ref{fig:res}(b)) leads also to the similar trend of the
damage parameter, see Fig.~\ref{fig:damage}(a). This can be
attributed to the lower exterior temperature and higher moisture
content in the surface layer caused by the driving-rain flux. It
is evident from the boundary conditions plotted in
Figs.~\ref{fig:bc}(a),(c). The calculated results promote the
capability of proposed governing equations to simulate a
degradation processes in the building materials exposed to real
weather conditions.
%
\subsection{Influence of porosity} \label{sec:InfPSD}

\begin{figure} [h!]
\begin{center}
\begin{tabular}{cc}
\includegraphics*[width=75mm,keepaspectratio]{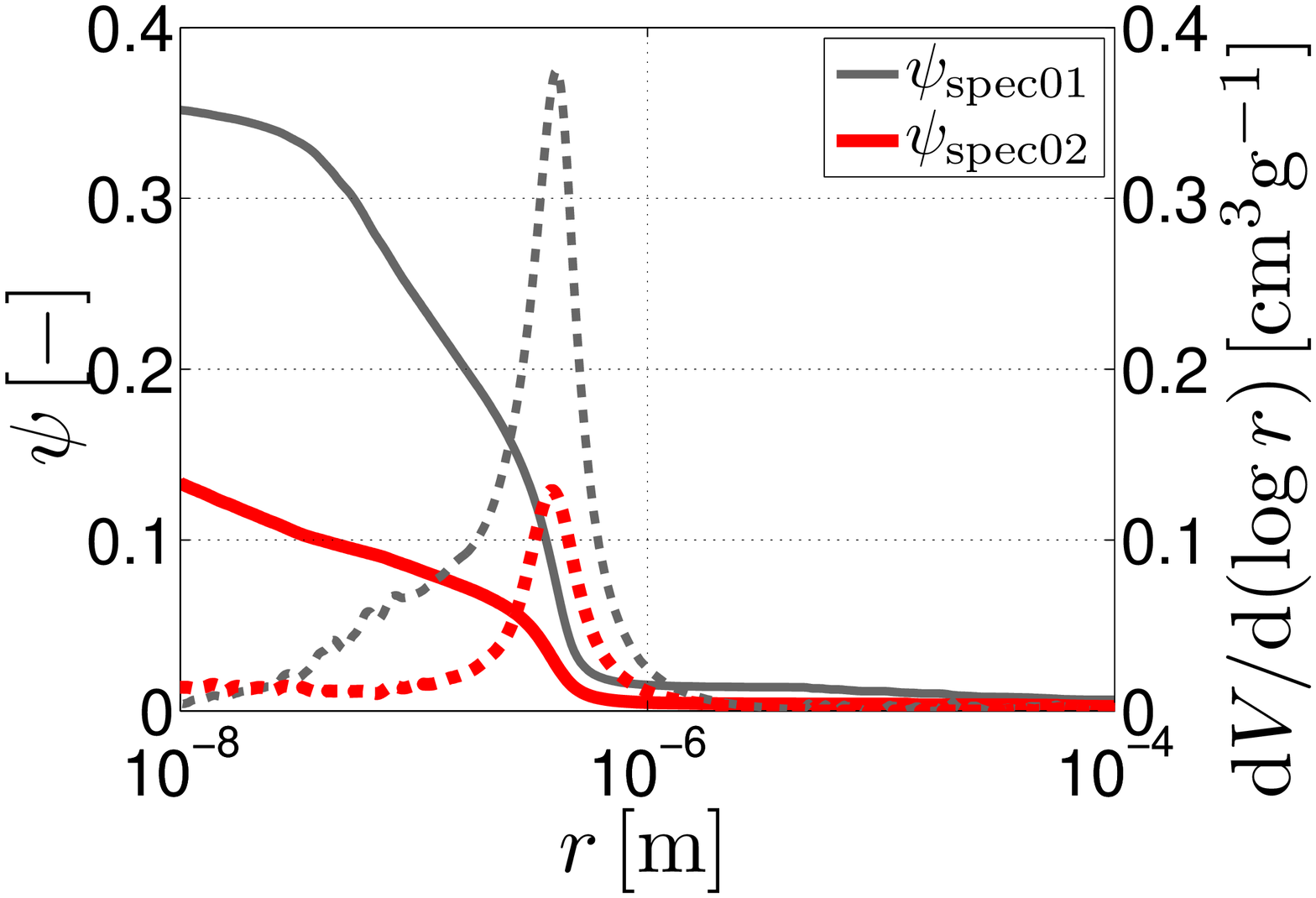}&
\includegraphics*[width=75mm,keepaspectratio]{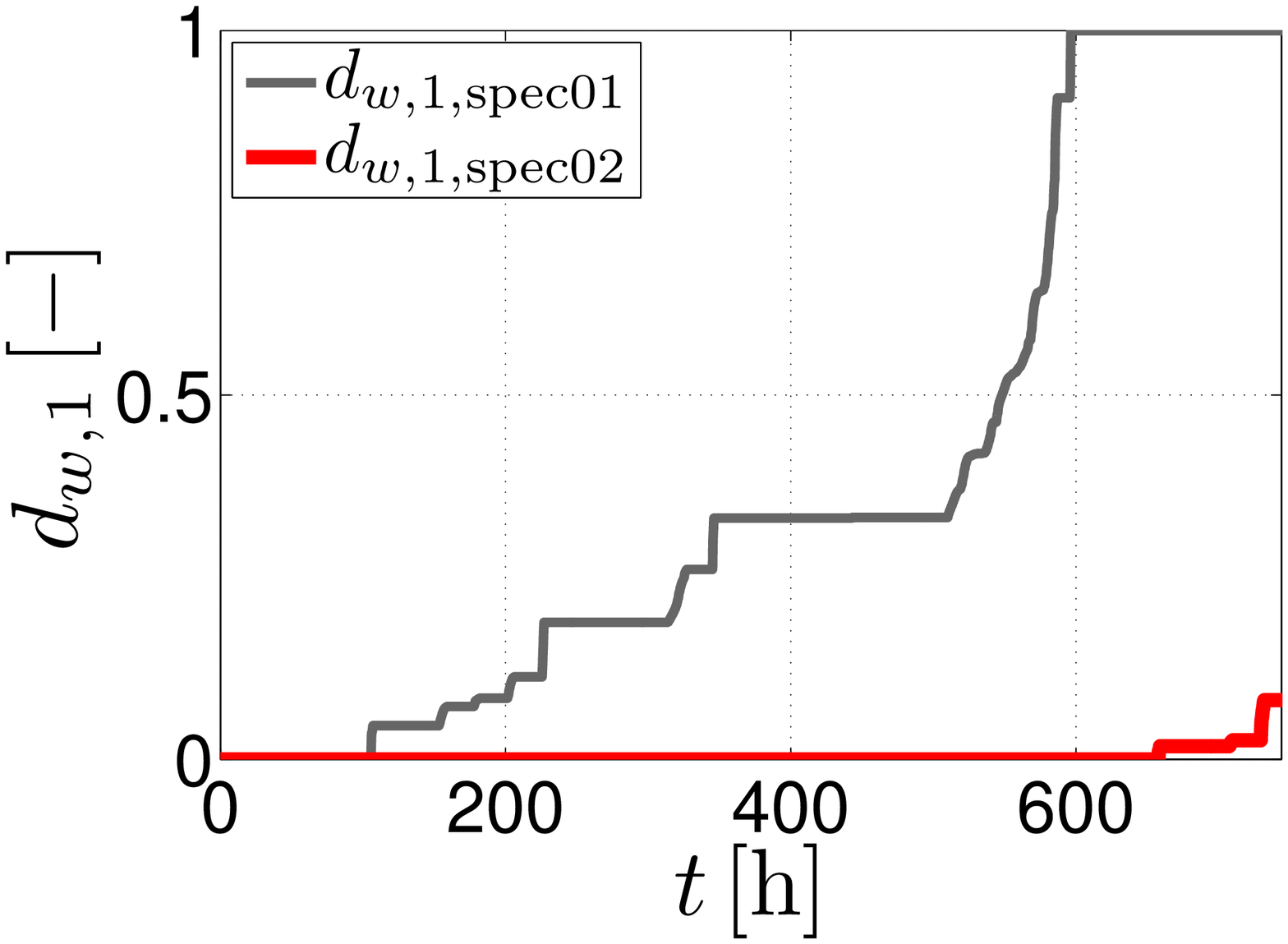}\\
(a)&(b) \\
\includegraphics*[width=75mm,keepaspectratio]{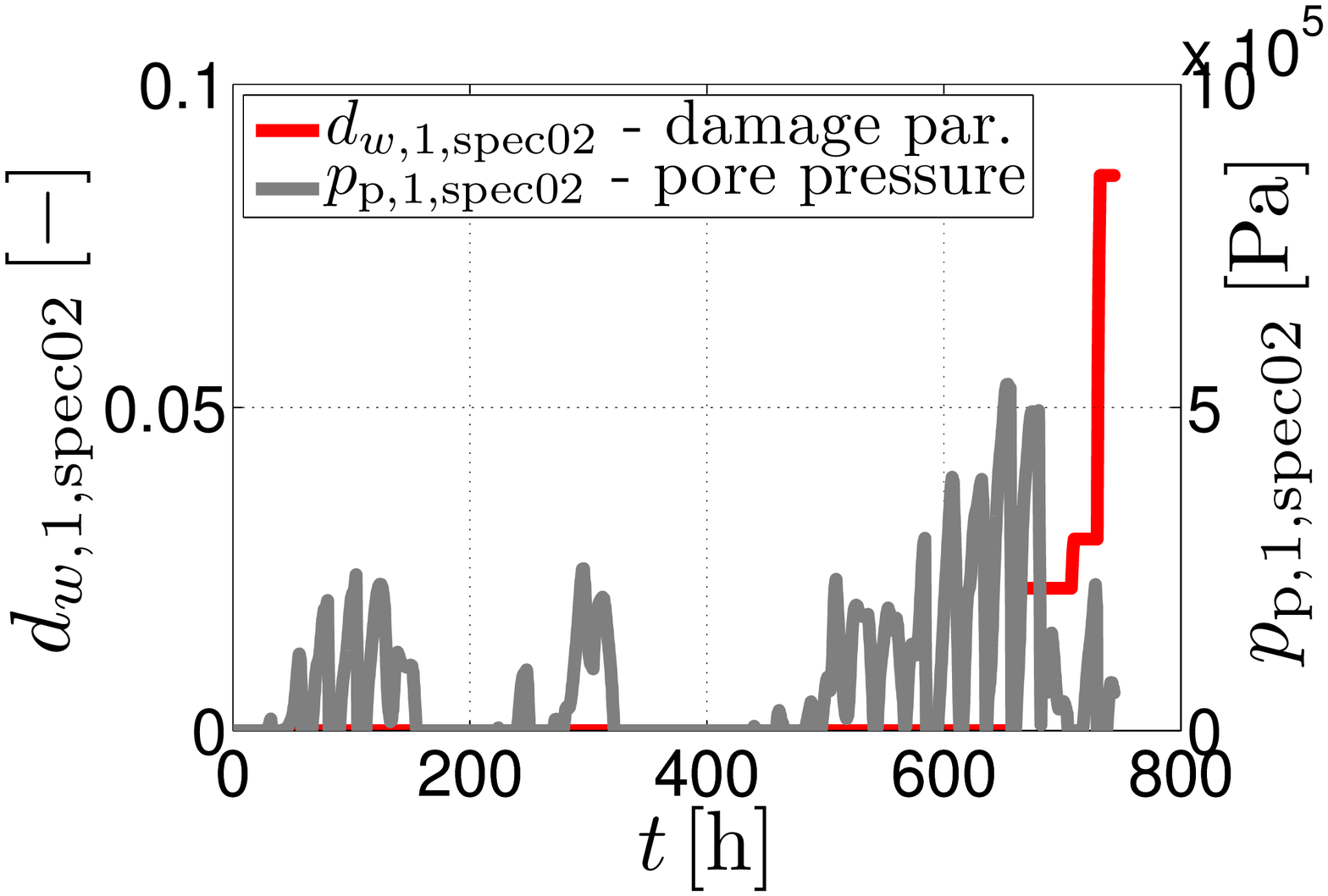}&
\includegraphics*[width=70mm,keepaspectratio]{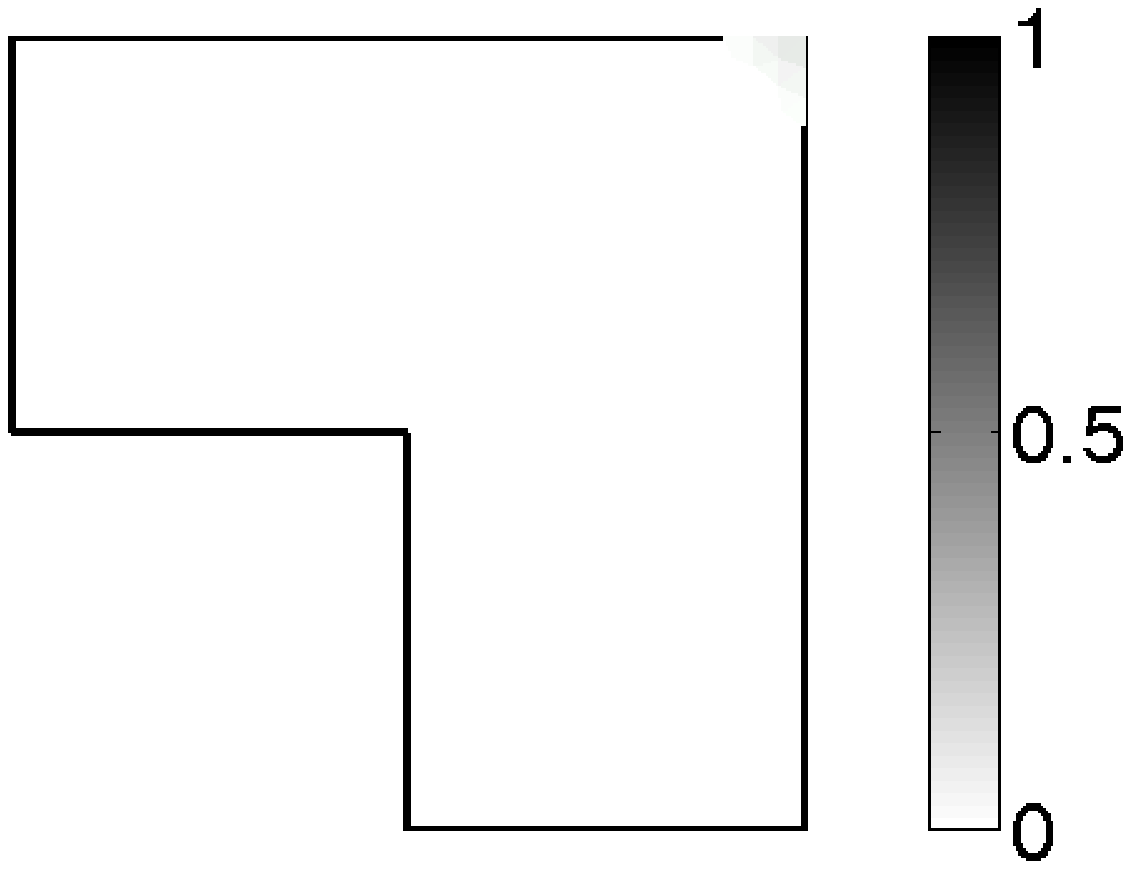}\\
(c)&(d) \\
\end{tabular}
\end{center}
\caption{(a) Two different pore size distributions and cumulative
porosity functions, (b) comparison of damage evolutions at node
$1$, (c) evolution of damage parameter
$d_{w,1,\mathrm{spec02}}\,\mathrm{[-]}$ and pore pressure
$p_{\mathrm{p},1,\mathrm{spec02}}\,\mathrm{[Pa]}$ at node $1$ for
porosity equal to $n_{\mathrm{spec02}}=0.13\,\mathrm{[-]}$, (d)
evolution of damage parameter
$d_{w,\mathrm{spec02}}\,\mathrm{[-]}$ at the end of analyzed time
period ($t=744\,\mathrm{[h]}$)} \label{fig:cmp}
\end{figure}

The formation of ice is mainly controlled by pore size
distribution, see ~\cite{Zuber:2000:CCR}. To address this issue we
consider the same input data as in the previous numerical example
except for the total porosity $n\,\mathrm{[-]}$ and cumulative
porosity $\psi\,\mathrm{[-]}$. Note that the structure of porous
system affects surely transport and mechanical properties of
mortars, but we focus here only on the influence of different
porosity to keep the numerical study clear and transparent, see
Eqs. (\ref{eq:icepress}) and (\ref{eq:linmom}). Therefore, two
different pore size distributions were taken into account, see
Fig.~\ref{fig:cmp}(a). Further we assume in calculations the
following values of total porosity
$\{n_{\mathrm{spec01}}=0.35\,\mathrm{[-]},n_{\mathrm{spec02}}=0.13\,\mathrm{[-]}\}$
representing material properties of the lime mortar and the lime
mortars with oil additive, respectively.

For a given pore size distributions, Figs.~\ref{fig:cmp}(b), (d)
shows the evolution of damage parameters in the critical location
of the $2$-D domain. It is shown that the value of total porosity
changes slightly, while the influence on the evolution of damage
parameter is significant. This is clearly observed from the
comparison of pore pressures displayed in Fig.~\ref{fig:damage}(a)
and Fig.~\ref{fig:cmp}(c). Combining all the previous results
suggests that the structure of porous system plays crucial role in
the resulting pore pressure and subsequently calculated internal
damage.

\section{Conclusions}
\label{sec:con}

This paper presents the numerical modeling of damage caused by ice
crystallization process in mortars. Attention is focused on the
thermo-hygro-mechanical model developed here in the framework of
uncoupled algorithmic scheme. Two particular issues were
addressed: (i) the formulation of material model based on laws of
the thermodynamics, poromechanics and damage mechanics, (ii)
influence of porosity on the mechanical behavior.

In particular, we employed K\"{u}nzel's model, which is
sufficiently robust to describe real-world materials, but which is
also highly nonlinear, time-dependent material model. Supported by
several successful applications in civil engineering we adopted
Biot's model and the nonlocal isotropic damage model in the
framework to simulate the frost action on porous media.

A crucial point in modeling of damage in mortars is the pore size
distribution. The obtained results suggest a high importance of
porosity on evolution of the damage parameter, at least for the
present material parameters and applied range of initial and
boundary conditions.

Finally a comparison of the numerical calculations with
experimental measurements is under current investigation and will
be presented elsewhere.

\vspace*{4mm}
\noindent
{\bf \large Acknowledgement}

\noindent This outcome has been achieved with the financial
support of project NAKI - DF11P01OVV0080 entitled
"High-performance and compatible lime mortars for extreme
application in restoration, repair and preventive maintenance of
architectural heritage". The experimental work performed at the
Institute of Theoretical and Applied Mechanics AS CR under the
leadership of Dr.~Zuzana Sl\'{i}\v{z}kov\'{a} is also gratefully
acknowledged.

\bibliography{references_rev}

\begin{thebibliography}{10}

\bibitem{Bazant:2002:JEM}
Z.~B. Ba\v{z}ant and J.~Jir\'{a}sek.
\newblock Nonlocal integral formulations of plasticity and damage: Survey of
  progress.
\newblock {\em Journal of Engineering Mechanics}, 128(11):1119--1149, 2002.

\bibitem{Biot:2002:JAP}
M.~A. Biot.
\newblock General theory of three-dimensional consolidation.
\newblock {\em Journal of Applied Physics}, 12(2):155--164, 1941.

\bibitem{Coussy:2007:CG}
O.~Coussy and P.~Monteiro.
\newblock Unsaturated poroelasticity for crystallization in pores.
\newblock {\em Computers and Geotechnics}, 34(4):279--290, 2007.

\bibitem{Coussy:2008:CCR}
O.~Coussy and P.~Monteiro.
\newblock Poroelastic model for concrete exposed to freezing temperatures.
\newblock {\em Cement and Concrete Research}, 38(1):40--48, 2008.

\bibitem{Fagerlund:1973:MS}
G.~Fagerlund.
\newblock Determination of pore-size distribution from freezing-point
  depression.
\newblock {\em Materials and Structures}, 6(3):215--225, 1973.

\bibitem{Gawin:2003:CMAME}
D.~Gawin, F.~Pesavento, and B.~A. Schrefler.
\newblock Modelling of hygro-thermal behaviour of concrete at high temperature
  with thermo-chemical and mechanical material degradation.
\newblock {\em Computer Methods in Applied Mechanics and Engineering},
  192(13):1731--1771, 2003.

\bibitem{Horak:2009}
M.~Hor\'{a}k.
\newblock Localization analysis of damage and plasticity models.
\newblock Master's thesis, CTU in Prague, 2009.
\newblock in Czech.

\bibitem{Kong:2011:EB}
F.~Kong and H.~Wang.
\newblock Heat and mass coupled transfer combined with freezing process in
  building materials: Modeling and experimental verification.
\newblock {\em Energy and Building}, 43(10):2850--2859, 2011.

\bibitem{Koster:2010:RILEM}
M.~Koster.
\newblock 3{D} multi-scale model of moisture transport in cement-based
  materials.
\newblock In W.~Brameshuber, editor, {\em International RILEM Conference on
  Material Science}, pages 165--176, 2010.

\bibitem{Kunzel:1996:IJHMT}
H.~M. K\"{u}nzel and K.~Kiessl.
\newblock Calculation of heat and moisture transfer in exposed building
  components.
\newblock {\em International Journal of Heat and Mass Transfer},
  40(1):159--167, 1996.

\bibitem{Kucerova:2013:AMC}
A.~Ku\v{c}erov\'{a} and J.~S\'{y}kora.
\newblock Uncertainty updating in the description of coupled heat and moisture
  transport in heterogeneous materials.
\newblock {\em Applied Mathematics and Computation}, 219(13):7252–--7261, 2013.

\bibitem{Lewis:1999}
R.~W. Lewis and B.~A. Schrefler.
\newblock {\em The Finite Element Method in the Static and Dynamic Deformation
  and Consolidation of Porous Media}.
\newblock John Wiley\&Sons, Chichester, England, 2nd edition, 1999.

\bibitem{Liu:2011:CRT}
L.~Liu, G.~Ye, E.~Schlagen, H.~Chen, Z.~Qian, W.~Sun, and K.~van Breugel.
\newblock Modeling of the internal damage of saturated cement paste due to ice
  crystallization pressure during freezing.
\newblock {\em Cement and Concrete Research}, 33(5):562--571, 2011.

\bibitem{Matala:1995}
S.~Matala.
\newblock {\em Effects of carbonation on the pore structure of granulated blast
  furnace slag concrete}.
\newblock PhD thesis, Helsinki University of Technology, 1995.

\bibitem{Milani:2012:CBM}
G.~Milani and A.~Tralli.
\newblock Simple lower bound limit analysis homogenization model for in- and
  out-of-plane loaded masonry walls.
\newblock {\em Construction and Building Materials}, 25(12):808--834, 2012.

\bibitem{Milani:2012:IJSS}
G.~Milani and A.~Tralli.
\newblock A simple meso-macro model based on {S}{Q}{P} for the non-linear
  analysis of masonry double curvature structures.
\newblock {\em International Journal of Solids and Structures},
  49(4):1999--2011, 2012.

\bibitem{Nunes:2012:EM}
C.~Nunes, Z.~Sl\'{i}\v{z}kov\'{a}, D.~K\v{r}i\'{a}nkov\'{a}, and
  D.~Frankeov\'{a}.
\newblock Effect of linseed oil on the mechanical properties of lime mortars.
\newblock In {\em Engineering Mechanics 2012}, 2012.

\bibitem{Multon:2012:IJNAMG}
A.~Sellier S.~Multon and B.~Perrin.
\newblock Numerical analysis of frost effects in porous media. {B}enefits and
  limits of the finite element poroelasticity formulation.
\newblock {\em International Journal for Numerical and Analytical Methods in
  Geomechanics}, 36(4):438--458, 2012.

\bibitem{Scherer:1993:JNS}
G.~W. Scherer.
\newblock Freezing gels.
\newblock {\em Journal of Non-Crystalline Solids}, 155(1):1--25, 1993.

\bibitem{Scherer:1999:CCR}
G.~W. Scherer.
\newblock Crystallization in pores.
\newblock {\em Cement and Concrete Research}, 29(8):1347–--1358, 1999.

\bibitem{Slizkova:2010:ROUD}
Z.~Sl\'{i}\v{z}kov\'{a}, M.~Drd\'{a}ck\'{y}, D.~Frankeov\'{a}, L.~Nos\'{a}l,
  and A.~Zeman.
\newblock Study of historic mortars from {C}harles {B}ridge.
\newblock In {\em Restaurov\'{a}n\'{i} a Ochrana Um\v{e}leck\'{y}ch d\v{e}l
  IV}, pages 26--28, 2010.
\newblock in Czech.

\bibitem{Sun:2010:CCR}
Z.~Sun and G.~W. Scherer.
\newblock Effect of air voids on salt scaling and internal freezing.
\newblock {\em Cement and Concrete Research}, 40(2):260--270, 2010.

\bibitem{Sykora:2012:JCAM}
J.~S\'{y}kora, T.~Krej\v{c}\'{i}, J.~Kruis, and M.~\v{S}ejnoha.
\newblock Computational homogenization of non-stationary transport processes in
  masonry structures.
\newblock {\em Journal of Computational and Applied Mathematics},
  18:4745–--4755, 2012.

\bibitem{Sykora:2013:AMC}
J.~S\'{y}kora, M.~\v{S}ejnoha, and J.~\v{S}ejnoha.
\newblock Homogenization of coupled heat and moisture transport in masonry
  structures including interfaces.
\newblock {\em Applied Mathematics and Computation}, 219(13):7275--–7285, 2013.

\bibitem{Tan:2011:CRT}
X.~Tan, W.~Chen, H.~Tian, and J.~Cao.
\newblock Water flow and heat transport including ice/water phase chase in
  porous media: Numerical simulation and application.
\newblock {\em Cold Regions and Technology}, 68(1--2):74--84, 2011.

\bibitem{Cerny:2002}
R.~\v{C}ern\'{y} and P.~Rovnan\'{i}kov\'{a}.
\newblock {\em Transport Processes in Concrete}.
\newblock London: Spon Press, 2002.

\bibitem{Wardeh:2008:CBM}
G.~Wardeh and B.~Perrin.
\newblock Numerical modelling of the behaviour of consolidated porous media
  exposed to frost action.
\newblock {\em Construction and Building Materials}, 22(4):600--–608, 2008.

\bibitem{Zuber:2000:CCR}
B.~Zuber and J.~Marchand.
\newblock Modeling the deterioration of hydrated cement systems exposed to
  frost action - {P}art 1: Description of the mathematical model.
\newblock {\em Cement and Concrete Research}, 30(12):1929--1939, 2000.

\end{thebibliography}

\end{document}